\documentclass{article}
\usepackage{graphicx}
\usepackage{amsmath,amssymb}
\usepackage{graphics,color,array,calc,rotating,epsfig,psfrag}
\numberwithin{equation}{section}
\usepackage{mathtools} 
\usepackage{amsmath}   
\usepackage{hyperref}
\usepackage{cite}
\usepackage{bm}
\usepackage{dcolumn}
\usepackage{float}
\usepackage{subfigure}
\usepackage{manyfoot}
\usepackage{epstopdf}
\usepackage{inputenc}
\usepackage{booktabs}  
\usepackage{siunitx}   
\usepackage{multirow}  
\DeclareNewFootnote{R}[roman]
\usepackage{amsfonts}
\usepackage[mathscr]{eucal}
\def\be{\begin{equation}} \def\ee{\end{equation}}
\def\bea{\begin{eqnarray}} \def\eea{\end{eqnarray}}

\newcommand{\eg}{{\it e.g.,}\ }

\newcommand{\nn}{\nonumber}
\begin{document}
\baselineskip 18pt%
\begin{titlepage}
\vspace*{1mm}%
\hfill%
\vspace*{15mm}%
\hfill
\vbox{
    \halign{#\hfil         \cr
          } 
      }  
\vspace*{20mm}

\begin{center}
{\large {\bf
 Scalar field perturbations in Non-commutative Schwarzschild spacetime
}}\\
\vspace*{5mm}
{  Majid Karimabadi\footnoteR{ma.karimabadi@hsu.ac.ir}, Davood Mahdavian Yekta\footnoteR{d.mahdavian@hsu.ac.ir}, S. A. Alavi\footnoteR{s.alavi@hsu.ac.ir}}\\

\vspace*{0.2cm}
{$^{}$ Department of Physics, Hakim Sabzevari University, P.O. Box 397, Sabzevar, Iran}\\
\vspace*{1cm}
\end{center}
\begin{abstract}
This work presents a comparative analysis of quasi-normal modes and 
ringdown profiles of scalar field perturbations in the non-commutative 
Schwarzschild black hole spacetime, focusing on two non-minimal 
curvature couplings: in the first, the scalar field is coupled to the Ricci
scalar of the background geometry, while in the second, its derivatives are coupled to the Einstein tensor. Using the sixth-order WKB approximation and the time-domain integration method, the dynamical stability is investigated across the parameter space. 
It is shown that while the tensor model loses its hyperbolic evolution for 
large couplings, no physical instability is found within its well-posed regime. 
Moreover, an opposite \(\ell\)-dependence of the threshold couplings is 
identified between the two models. Finally, using the result of the scalar sector, a 
model-dependent length scale for the non-commutative parameter is estimated 
in the case of light primordial black holes.
\end{abstract}

\end{titlepage}

\section*{Introduction}

The advent of gravitational wave astronomy, marked by the historic detection of binary black hole mergers \cite{LIGOScientific:2016aoc,LIGOScientific:2016sjg}, has opened a new window for studying strong gravitational phenomena. A key observable in such events is the ringdown phase, during which the merged black hole settles into a stationary state by emitting gravitational radiation characterized by a discrete set of damped oscillations known as quasi-normal modes (QNMs). Black holes are constantly in perturbed states due to the interactions with other black holes, fields, or neutron stars. QNMs are determined by the black hole parameters and the underlying theoretical model (for reviews on QNMs, see \cite{Kokkotas:1999bd,Konoplya:2011qq}).

Not only the QNMs provide a test bed for general relativity in the strong-field regime, but they also serve as probes for modifications of gravity or matter couplings. Horndeski gravity \cite{Horndeski:1974wa} is the most well-known theory in modified gravity. A subclass of the Horndeski Lagrangian that includes non-minimal interactions between scalar fields and curvature tensors has received considerable attention due to its distinctive phenomenological implications \cite{Clifton:2011jh}. This includes inflationary models with a slow-roll phase \cite{Amendola:1993uh,Sushkov:2009hk,Artymowski:2012is}, the redshift behavior of perturbations during inflation \cite{Papantonopoulos:2019eff,Dalianis:2019vit}, and cosmological models of dark matter \cite{Abreu:1994fd,Mannheim:2005bfa,Saridakis:2010mf}.

On the other hand, non-commutative (NC) geometry has been proposed as a potential avenue for incorporating quantum gravitational corrections into classical spacetime \cite{Ahluwalia:1993dd,Doplicher:1994zv}. Inspired by string theory \cite{Seiberg:1999vs,Ardalan:1998ce} and non-perturbative formulations of quantum gravity \cite{Kempf:1994su,AmelinoCamelia:2008qg}, NC geometry suggests that the coordinates of spacetime may not commute at high energy scales. There are many different methods specifically devised for dealing with the deformed structures of spacetime \cite{Douglas:2001ba,Szabo:2001kg,Aschieri:2005zs,connes2019}. In this paper, we use the coordinate coherent state formalism \cite{Snyder:1946qz,Smailagic:2003yb,Smailagic:2003rp}. This approach leads to regular black hole solutions \cite{Nicolini:2005vd,Ansoldi:2006vg}. QNMs contain intrinsic information about black hole parameters and also carry information about the properties of the underlying spacetime structure, i.e., deviations of the spacetime structure from the usual notion of a smooth spacetime manifold should be imprinted in the QNM spectra.

Our main motivation in this paper is to investigate the QNMs of perturbations of a scalar field, non-minimally coupled to curvature tensors, in the background of a non-commutative Schwarzschild (NC-Sch) black hole. We analyze two different scenarios. In the \textit{scalar-coupled model}, the scalar field is coupled non-minimally to the Ricci scalar $R$ of the black hole geometry. The field equation reduces to a Klein-Gordon equation with a modified mass term, which introduces curvature-dependent corrections to the dynamics of the probe field. In the \textit{tensor-coupled model}, the field derivatives are non-minimally coupled to the Einstein tensor $G_{\mu\nu}$ of the geometry. It should be noted that in both scenarios, the NC-Sch black hole background remains unaffected by the interaction with the real scalar field, i.e., we treat the scalar field as a test field and neglect backreaction on the geometry.

The scalar field perturbations in both models are governed by a master equation known as the Regge-Wheeler equation \cite{Regge:1957td}, which is transformed into a Schr\"{o}dinger-like equation by a simple tortoise coordinate transformation \cite{Zerilli:1970se,Teukolsky:1972my}. In general, the QNMs are found by solving this equation analytically; however, analytic treatment is only possible for a highly restricted range of parameters. There exist a number of methods for solving this equation semi-analytically and numerically \cite{Ferrari:1984zz,Leaver:1985ax,Motl:2002hd,Saad:2003vhv}. The most widely used among them is the Wentzel-Kramers-Brillouin (WKB) approximation. The lowest-order approximation for Schwarzschild black holes was considered in \cite{Schutz:1985km}, and its extensions to third order were accomplished in \cite{Iyer:1986np,Iyer:1986nq,Kokkotas:1988fm,Kokkotas:1991vz}. The WKB approach has also been extended to higher orders in Refs.~\cite{Konoplya:2003ii,Konoplya:2019hlu,Matyjasek:2017psv}.

We will compute the QNMs using the sixth-order WKB approximation, which is a popular choice due to its high accuracy. We investigate how the NC parameter and the curvature coupling constants affect the real and imaginary parts of the QNM frequencies in the two scenarios. The QNMs of scalar field perturbations with non-minimal couplings for various black hole backgrounds other than NC-Sch have been investigated in Refs.~\cite{Chan:1996yk,Aros:2002te,Kanti:2014dxa,Rincon:2018ktz,MahdavianYekta:2019pol,MahdavianYekta:2018lnp} for the scalar-coupled model, and in Refs.~\cite{Chen:2010ru,Chen:2010qf,Yao:2011kf,Minamitsuji:2014hha,Yu:2018zqd,Yan:2020nvk} for the tensor-coupled model. Additionally, the dynamical evolution of a scalar field in NC inspired black holes with minimal coupling has been studied in Refs.~\cite{Panotopoulos:2019gtn,Campos:2021sff,Heidari:2023egu,Yan:2023pxj,Batic:2024zbx,Ma:2024pyc}, however, in this paper we decide to bring forward a detailed comparison between the two models on the same background and investigate the threshold couplings and NC parameter bound based on the onset of instability. In particular, though the instability for minimally coupled perturbations on this background have initially been discussed in Ref.~\cite{Batic:2024zbx}, our work presents further the instabilities induced by the non-minimally curvature couplings as well.

The other main motivation in this paper is to study the effects of physical parameters in the models on the time-domain ringdown profiles of scalar field propagating in the vicinity of NC-Sch black hole, using the time-domain integration method proposed in Ref.~\cite{Gundlach:1993tn}. Ringdown profiles are particularly important because they encode not only the characteristic QNM frequencies of the black hole, but also the stability properties of the perturbed system. Several related works have followed this method to investigate the ringdown process through black hole perturbations \cite{Abdalla:2006vb,Chirenti:2007mk,Toshmatov:2017bpx,Konoplya:2018qov,Abdalla:2018ggo,Chakrabarty:2018skk,Berti:2022xfj,DuttaRoy:2022ytr,Redondo-Yuste:2023ipg}. In this respect, the threshold value of the coupling constant plays an important role in determining a transition point between stable and unstable configurations. This critical value is therefore essential not only for understanding the onset of instability in non-minimally coupled theories but also for constraining allowed values of the coupling constants from black hole perturbation observations. Using the implication of ringdown profiles together with the critical values of the coupling we investigate the stability in two models and, moreover, determine a model-dependent length scale for the NC parameter -- as the third motivation in this study. 

The structure of the paper is as follows. In Sec.~\ref{sec1}, we review the construction of NC-Sch black holes using the coordinate coherent state approach. In Secs.~\ref{sec2} and~\ref{sec3}, we introduce the two non-minimal coupling models: the scalar and tensor models respectively. For each model, we investigate the dynamical perturbations of a scalar field in the NC-Sch background and compute the QNM frequencies using the sixth-order WKB method, followed by a comparative analysis of the two scenarios. The time-domain ringdown profiles of the dynamical perturbations are discussed in Sec.~\ref{sec4}. In this section, we also study the threshold values and find the length scale for the NC parameter.

\section{NC geometry in coordinate coherent state formalism} \label{sec1}

The incorporation of NC geometry into gravitational theories results in significant modifications to classical black hole solutions. One of the most widely used approaches to implement noncommutativity is the \textit{coordinate coherent state formalism} \cite{Smailagic:2003yb,Smailagic:2003rp}. In this framework, the effects of noncommutativity are implemented by replacing the classical point-like mass and charge distributions with smeared Gaussian profiles, there by avoiding the curvature singularity  of classical solutions. This method does not deform the spacetime manifold itself, but rather modifies the matter source, preserving general covariance.
In this formalism, the standard Schwarzschild  black hole metrics receive NC corrections that regularize the curvature at short distances \cite{Nicolini:2005vd}, but it remains spherically symmetric and static
\be\label{bm}
ds^2 = -f(r)\,dt^2 + \frac{dr^2}{f(r)} + r^2 \left(d\vartheta^2 + \sin^2\vartheta\,d\varphi^2\right),
\ee
where the lapse function $f(r)$ is modified due to the NC effects. For the NC-Sch black hole, it takes the form
\be\label{Sch}
f(r) = 1 - \frac{4M}{r\sqrt{\pi}} \gamma\left(\frac{3}{2}, \frac{r^2}{4\theta}\right).
\ee
$\gamma(a, x)$ is the incomplete Gamma function where via $\gamma(a, x)+\Gamma(a, x)=\Gamma(a)$, one has
\be
\gamma\left(\frac{3}{2}, \frac{r^2}{4\theta}\right)=\frac{\sqrt{\pi}}{2}-\Gamma\left(\frac{3}{2}, \frac{r^2}{4\theta}\right),
\ee
such that the Schwarzschild blackening factor changes to
\be \label{schb}
f(r) = 1 - \frac{2M}{r}+\frac{4M}{r\sqrt{\pi}}\Gamma\left(\frac{3}{2}, \frac{r^2}{4\theta}\right).
\ee

This kind of smeared mass distribution effectively removes the curvature singularity at $r = 0$, making the resulting spacetime regular everywhere. It can be verified by examining the curvature invariants near the origin. For the NC-Sch black hole (\ref{Sch}), the curvature invariants behave as
\bea\label{invs}
R_{\mu\nu\rho\lambda}R^{\mu\nu\rho\lambda} &\xrightarrow{r \to 0}& \frac{8M^2}{3\pi \theta^3}, \nn \\
R_{\mu\nu}R^{\mu\nu} &\xrightarrow{r \to 0}& \frac{4M^2}{\pi \theta^3}, \\
R &\xrightarrow{r \to 0}& \frac{4M}{\sqrt{\pi} \theta^{3/2}}, \nn
\eea
which are finite for all $r$ as long as $\theta > 0$. This regularity for the quantities in (\ref{invs}) is one of the key advantages of employing the NC geometry: the naked singularity at $r = 0$ is replaced by a de Sitter-like core. Moreover, the behavior of the solutions in the  limit $\theta \rightarrow 0$ reproduces the classical results. Specifically, the function $f(r)$ in (\ref{Sch})  asymptotically approaches the familiar Schwarzschild  metric. This correspondence ensures consistency with general relativity at large distances or low energies, where NC effects become negligible.

The existence of event horizon in this geometry depends strictly on the values of mass $M$ and NC parameter $\theta$. For sufficiently small mass or large $\theta$, the lapse function $f(r)$ may not vanish for any positive $r$, indicating the absence of any horizon. In such cases, the solution represents a regular horizonless object rather than a black hole. For the NC-Sch black hole, it has been shown that the condition for the existence of horizon(s) is  \cite{Nicolini:2005vd,Karimabadi:2018sev}
\be \label{ran}
\theta \leq \left(\frac{M}{1.9}\right)^2.
\ee
Therefore, our analysis is confined to the parameter range $0 \leq \theta \leq \left(\frac{M}{1.9}\right)^2$ to ensure that at least there is an event horizon (like an extremal black hole). More mathematical details about the NC-Sch spacetime and its higher dimensional counterparts along with their effects on the gravitational measurements can be found in Refs.~\cite{Karimabadi:2018sev,Yekta:2019wlw,Filho:2022zdh,Heidari:2023bww}.
\section{First scenario: Non-minimal coupling to Ricci scalar}\label{sec2}

In the following we investigate the dynamics of a scalar field which is non-minimally coupled to the Ricci scalar of the NC-Sch black hole alluded in the previous section. The gravitational action describing this kind of interactive theory is given by
\be \label{sact}
S = -\frac{1}{2} \int d^4x\, \sqrt{-g} \left( g^{\mu\nu} \nabla_\mu \phi \nabla_\nu \phi + \mu^2 \phi^2 + \zeta R \phi^2 \right),
\ee
where $R$ is the Ricci scalar associated with the background geometry and $\zeta$ defines the corresponding coupling (a dimensionless constant) to the scalar field $\phi$ of mass $\mu$. The metric $g^{\mu\nu}$ is the inverse of NC-Sch metric given by Eq.~(\ref{bm}). The case with $\mu = 0$ and $\zeta = \frac{1}{6}$ corresponds to a theory of conformal coupling \cite{Mannheim:2011ds} which means the  theory becomes invariant under local conformal transformations of the metric. One of the motivations of such conformally coupled theories is that to explain dark matter, dark energy and aspects of quantum gravity \cite{Abreu:1994fd,Mannheim:2005bfa}.

By varying the action (\ref{sact}) with respect to $\phi$, we obtain a modified Klein-Gordon equation as follows
\be \label{eomsc}
\left[ \Box - \mu^2 - \zeta R \right] \phi(t, r, \vartheta, \varphi) = 0,
\ee
where $\Box \equiv \nabla_\mu \nabla^\mu$ is the covariant d'Alembertian operator. This equation describes the dynamics of scalar field in the background of NC-Sch black hole.
In this paper, we treat the scalar field as a test field which does not backreact on the geometry, however, in any real physical interaction of a field in the vicinity of a black hole, it can never be represented exactly as a pure test field, even when the coupling terms might be neglected for practical reasons.

In order to study the perturbations of scalar field via master equation (\ref{eomsc}), we use an ansatz with separation of variables as follows
\be \label{ansatz}
\phi(t, r, \vartheta, \varphi) = A(r) Y_{\ell m}(\vartheta, \varphi) e^{-i \omega t},
\ee
where $Y_{\ell m}(\vartheta, \varphi)$ are the scalar spherical harmonics, satisfying $\Box_{(\vartheta,\varphi)} Y_{\ell m} = -\ell(\ell+1) Y_{\ell m}$, and $\omega$ is a complex variable representing the QNM frequency. In fact, the time dependent term $e^{-i \omega t}$ reflects the dissipative nature of QNMs, that is if one cast the frequency as $\omega = \omega_R + i \omega_I$ then the stability of field near the geometry requires $\omega_I < 0$ to ensure decaying perturbations, while $\omega_I > 0$ leads to exponential growth and signals instability. In general, the master equation of scalar field can be rewritten as a Schr\"{o}dinger-like equation. For Eq.~(\ref{eomsc}) this is achieved by redefinition $A(r) = \frac{\psi(r)}{r}$ where the power of $r$ in the denominator depends on the dimension of spacetime, then
\be \label{reqs}
- f(r) \frac{d}{dr} \left[ f(r) \frac{d}{dr} \psi(r) \right] + V(r) \psi(r) = \omega^2 \psi(r).
\ee

Following the standard tortoise coordinate transformation by
\be \label{tc}
\frac{dr^*}{dr} = \frac{1}{f(r)},
\ee
one can recast Eq.~(\ref{req}) into the standard Regge-Wheeler wave equation \cite{Regge:1957td}:
\be \label{sceq}
\frac{d^2 \psi}{d{r^*}^2} + \left[ \omega^2 - V(r) \right] \psi = 0,
\ee
where $V(r)$ is an effective potential encapsulating the effects of spacetime geometry and interactions which is given by
\be \label{veff}
V(r) = f(r) \left( \mu^2 + \frac{\ell(\ell+1)}{r^2} + \frac{f'(r)}{r} + \zeta R \right).
\ee
The term \(\frac{f'(r)}{r}\) accounts for a gravitational curvature correction to the potential, while \({\ell(\ell+1)}/{r^2}\) represents the centrifugal barrier arising from angular momentum. The term \(\zeta R\) encodes the contribution from the curvature coupling, and \(\mu^2\) reflects the influence of the scalar field's mass on the wave evolution.

In order to determine the oscillation modes of NC-Sch black hole, which correspond
to natural solutions of Eq.~(\ref{sceq}), we must impose physically appropriate
boundary conditions at the horizon (\(r^* \to -\infty\)) and at spatial infinity (\(r^* \to +\infty\)); purely ingoing waves at the horizon and purely outgoing waves at spatial infinity. These conditions define the QNM spectra, which carries characteristic signatures of both the background geometry and the nature of the perturbing field (for more about boundary conditions see \eg \cite{Berti:2009kk}).

\begin{figure}[H]
\centering
\subfigure[$\mu=0.5$, $\zeta=50$, $\ell=10$]
{\includegraphics[width=.47\textwidth]{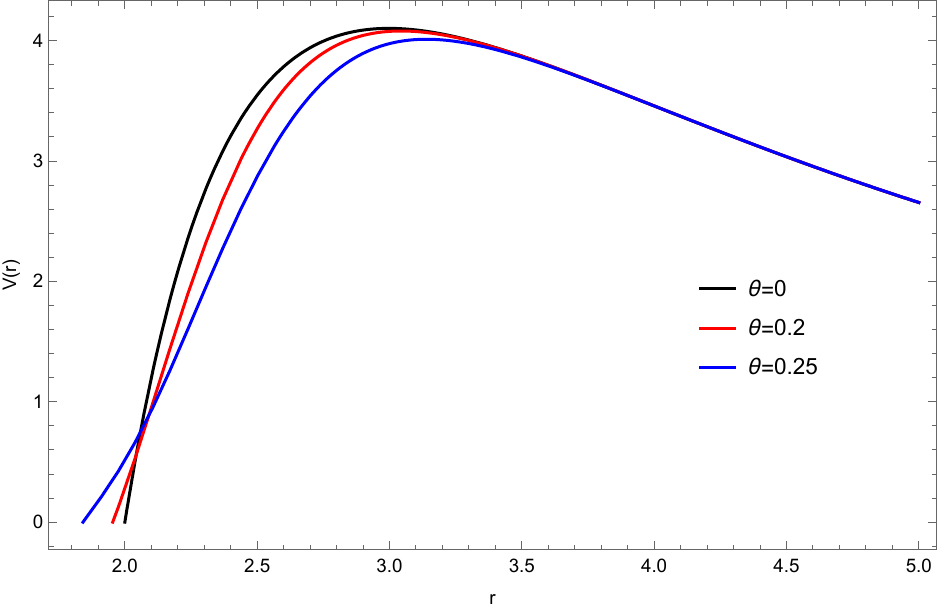}}
\subfigure[$\mu=0.5$, $\theta=0.2$, $\ell=10$]
{\includegraphics[width=.47\textwidth]{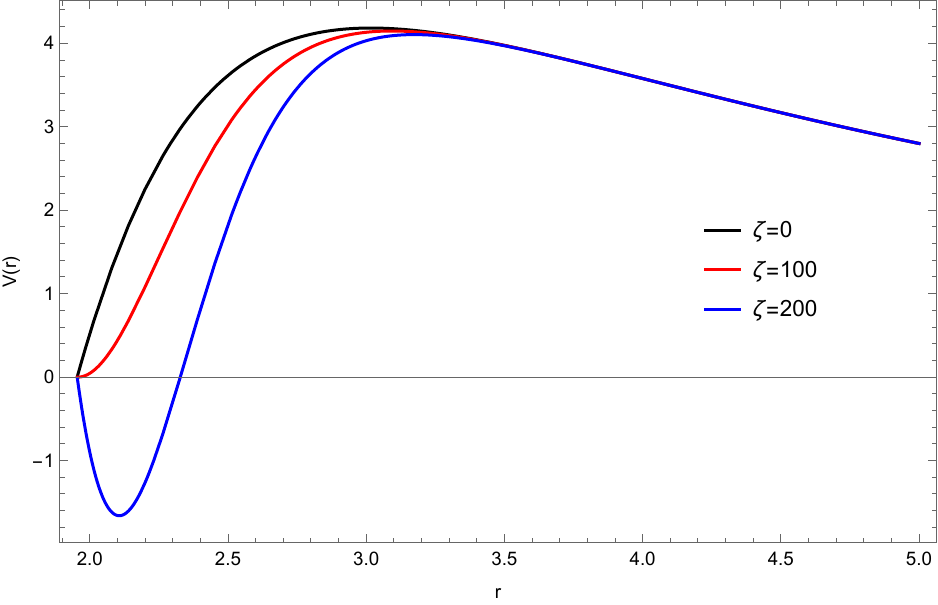}}
\subfigure[$\zeta=50$, $\theta=0.2$, $\ell=10$]
{\includegraphics[width=.47\textwidth]{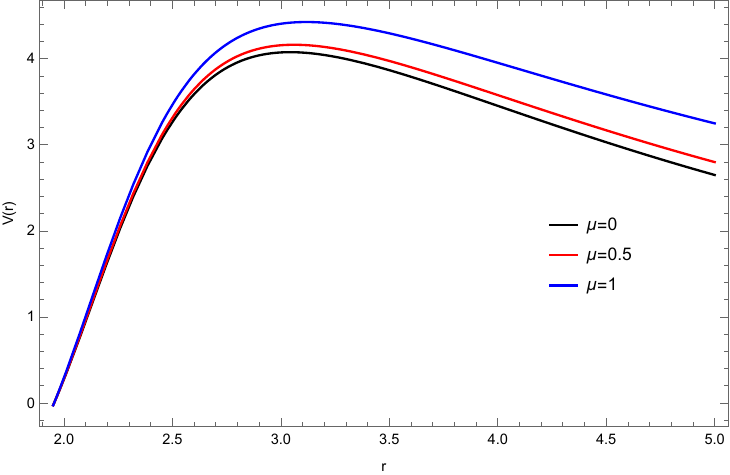}}
\subfigure[$\mu=0.5$, $\theta=0.2$, $\zeta=50$]
{\includegraphics[width=.47\textwidth]{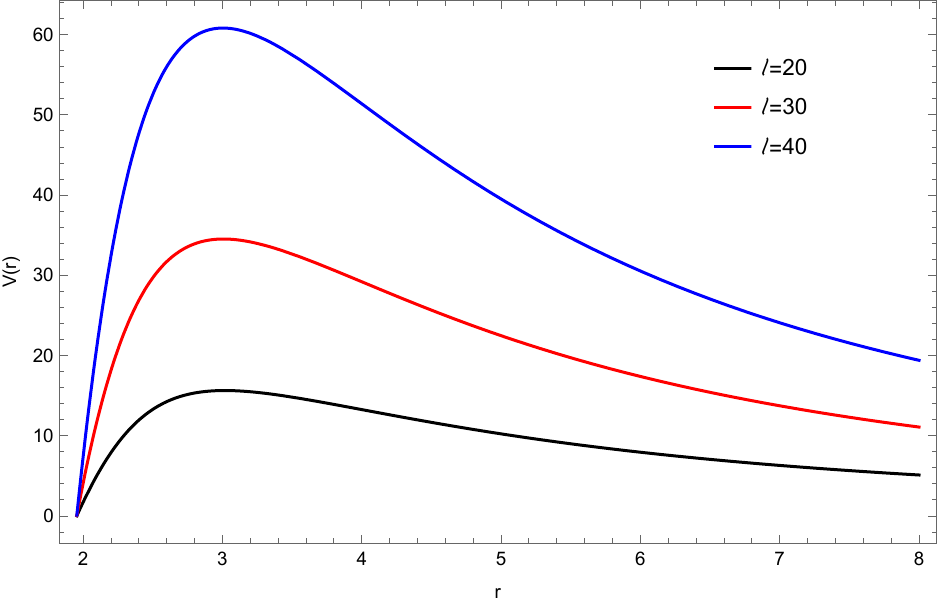}}
\caption{Effective potential $V(r)$ for different $\theta$,  $\zeta$, $\mu$ and $\ell$ }
\label{f1}
\end{figure}

The behavior of the effective potential $V(r)$ as a function of the radial coordinate $r$ for different values of the parameter space is plotted in Fig.~(\ref{f1}). Hereafter, without loss of generality and for simplicity, we set the black hole mass to $M = 1$. In panel (a), increasing the NC parameter $\theta$ within the range specified in Eq.~\eqref{ran} leads to a reduction in the potential barrier and a slight outward shift of its peak. This reflects the smoothing effect of NC geometry, where the gravitational source is smeared over a minimal length scale. These modifications, in turn, can influence the QNM spectrum in both the real and imaginary parts of the perturbation frequencies. For instance, the reduction in the oscillatory mode is consistent with a lower potential peak that is easier to overcome.

Panel (b) shows that a distinct structural transformation occurs as the coupling parameter $\zeta$ increases. Specifically, a negative potential well forms near the black hole horizon, situated just before the primary potential barrier. This new feature significantly modifies the shape of the effective potential by introducing more than two classical turning points, thereby invalidating the standard WKB approximation, which assumes a single barrier with exactly two such points. In these cases, alternative numerical methods, such as shooting techniques or time-domain integration, are necessary to accurately compute the QNM spectrum. Moreover, the plot reveals that increasing $\zeta$ leads to the suppression of the potential peak, indicating enhanced dissipation in the system.

As shown in panel (c), increasing the field mass $\mu$ raises the height of the potential barrier, as expected, since enhancing the overall energy content of the system results in stronger distortions of spacetime. Consequently, the QNM characteristics are modified: heavier fields tend to produce modes with higher oscillation frequencies and reduced damping, leading to longer-lived perturbations, especially at larger values of $\mu$ \cite{Ohashi:2004wr,Konoplya:2004wg}. Finally, panel (d) illustrates that as the angular momentum quantum number $\ell$ increases, the potential barrier also increases due to the enhanced centrifugal term $\ell(\ell+1)/r^2$, which acts as a repulsive force. Additionally, the peak of the potential shifts inward, moving slightly away from the photon sphere located at $r = 3M$ \cite{Synge:1966okc}. These changes lead to an increase in the oscillation frequency of the QNMs and a reduction in the damping rate, resulting in more sharply defined and longer-lived modes.

 \subsection{QNMs in the Scalar model}

In this section, we investigate the quasi-normal frequencies of perturbations of a scalar field in the background of NC-Sch black hole (\ref{Sch}) in the first scenario. Following \cite{Konoplya:2003ii,Matyjasek:2017psv}, we use the sixth-order WKB approximation to compute QNMs, which is used for solving wave equations with potential barriers in black hole perturbation theory. This method is efficient when the effective potential has a single, well-defined peak and asymptotes to a constant values at large distances such that the horizon boundary conditions are satisfied. Up to order \(N\), the WKB formula approximates the QNMs by expanding around the local maximum of the effective potential \(V(r^*)\), where \(r^*\) is defined by Eq. (\ref{tc}). In order to validation of the QNM computed from WKB approach one can also extract them by using a semi-analytical spectral method expanded into areas beyond the reach of the WKB approximation and would be useful for carefully studying the instability regime from the time-domain results. A number of perturbative analyses by using the spectral method can be found in Refs.~\cite{Batic:2024vwm,Batic:2024vsb,Batic:2025bkk,Batic:2025xge,Batic:2025hgp,Batic:2026jdt}.

The WKB method up to sixth order is given by the following relation
\be \label{w6}
\frac{i(\omega^2 - V_0)}{\sqrt{-2 V_0''}} - \sum_{j=2}^{6} \Lambda_j = n + \frac{1}{2}, \quad n = 0, 1, 2,... ,
\ee
where \(\omega\) is the complex QNM frequency, \(V_0\) is the value of the effective potential at its maximum, while \(V_0''\) is its second derivative with respect to \(r^*\) at that point. The terms \(\Lambda_j\) are higher-order corrections at order \(j\), which depend on the higher derivatives of \(V_0\) and the overtone number \(n\) (Their expressions are given in Ref.~\cite{Konoplya:2003ii}).
The real part \(\text{Re}[\omega]\) corresponds to the oscillation frequency of the mode, while the imaginary part \(\text{Im}[\omega]\) represents its damping rate due to the black hole absorption, tunneling and coupling constant. In general, the WKB formula (\ref{w6}) is most reliable when the multipolar number $\ell$ is significantly larger than the overtone number \( n \) while its accuracy decreases for highly damped modes with large \(n\), where the method becomes less effective. On the other hand, as shown in Fig.~(\ref{f1}d), for sufficiently large $\ell$ the potential barrier remains well-defined, and the WKB approximation holds with good accuracy.

\begin{figure}[H]
\centering
{\includegraphics[width=.49\textwidth]{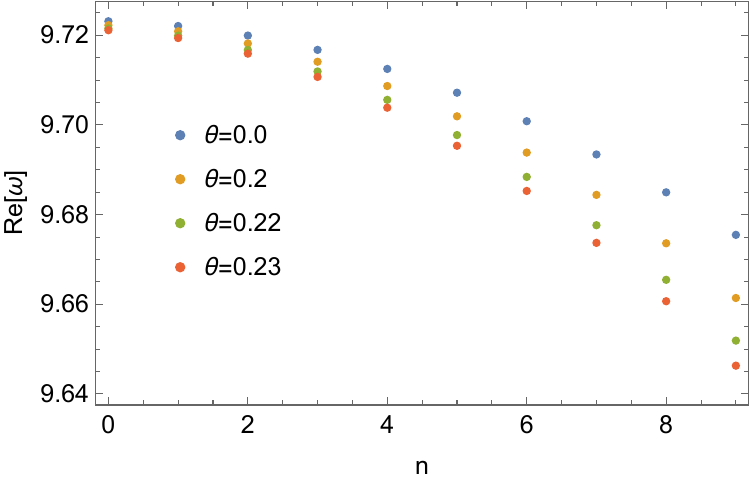}}
{\includegraphics[width=.49\textwidth]{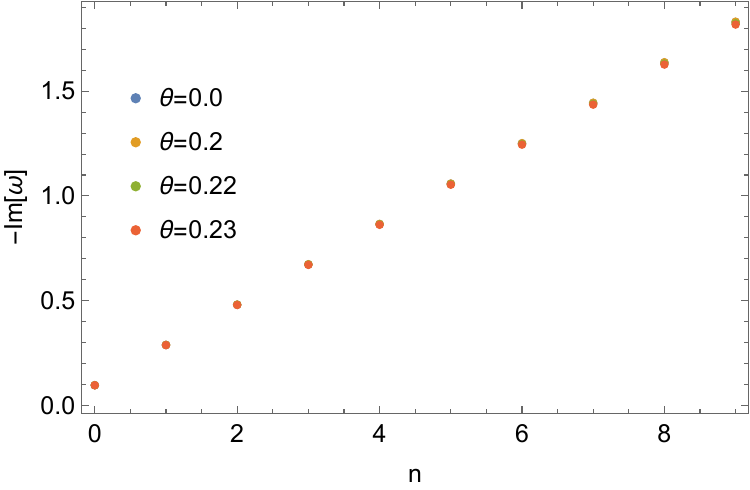}}
\caption{Variation of the real and imaginary parts of QNMs vs $n$ for different values of $\theta$ when $\mu = 0.5$, $\zeta = 50$, and $\ell = 50$.  }
\label{f2}
\end{figure}

\begin{figure}[H]
\centering

{\includegraphics[width=.49\textwidth]{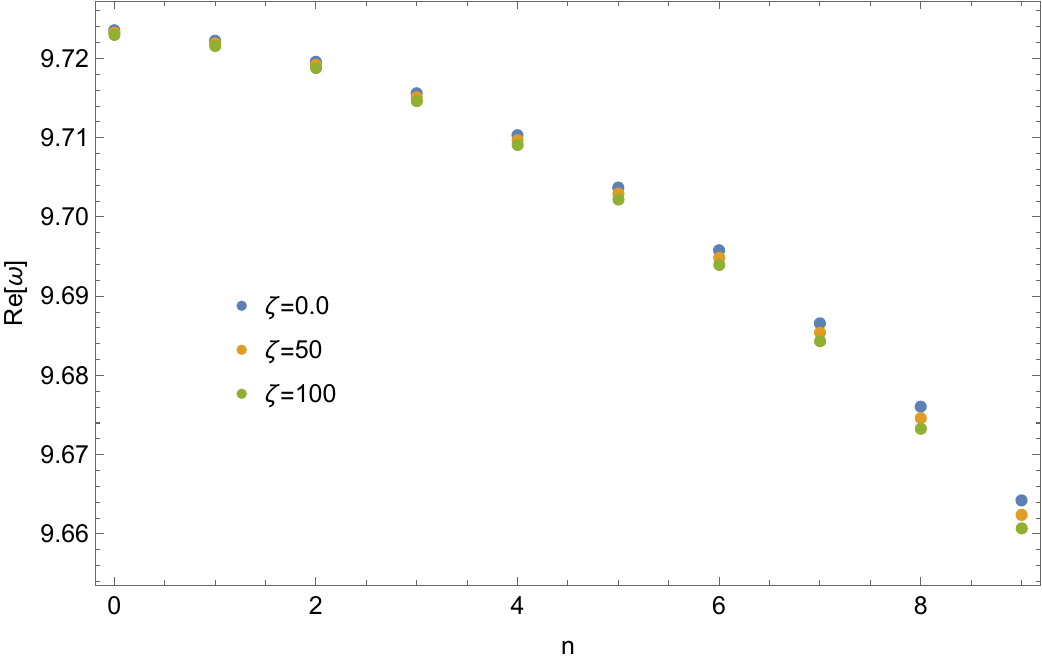}}
{\includegraphics[width=.49\textwidth]{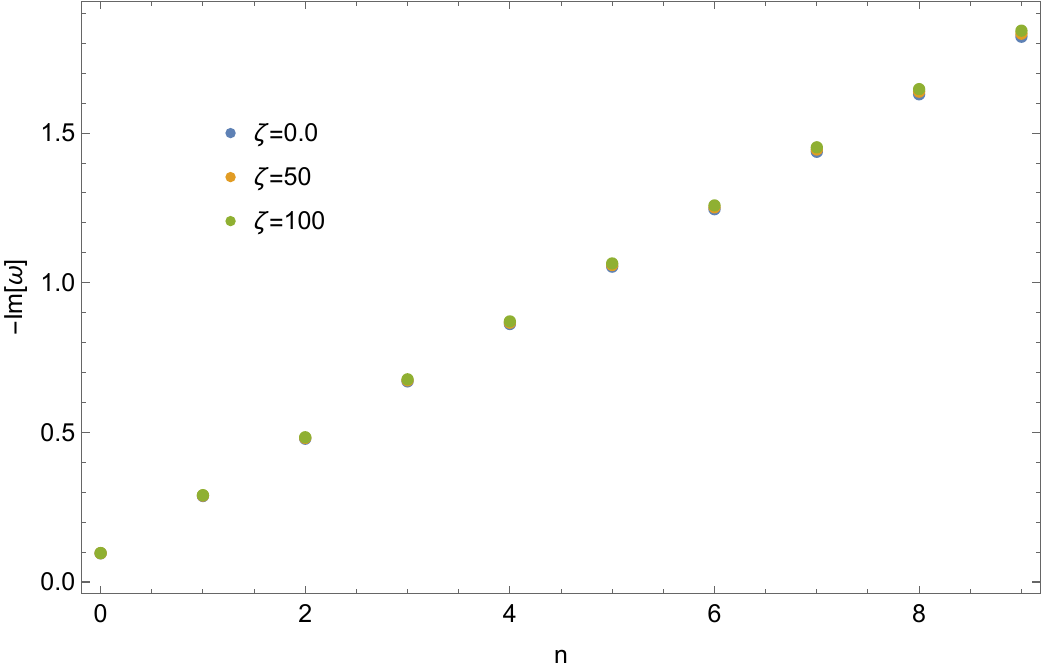}}
\caption{Real and imaginary parts of QNMs vs $n$ for different values of $\zeta$ when $\mu = 0.5$, $\theta = 0.2$, and $\ell = 50$.  }
\label{f3}
\end{figure}

\begin{figure}[H]
\centering

{\includegraphics[width=.49\textwidth]{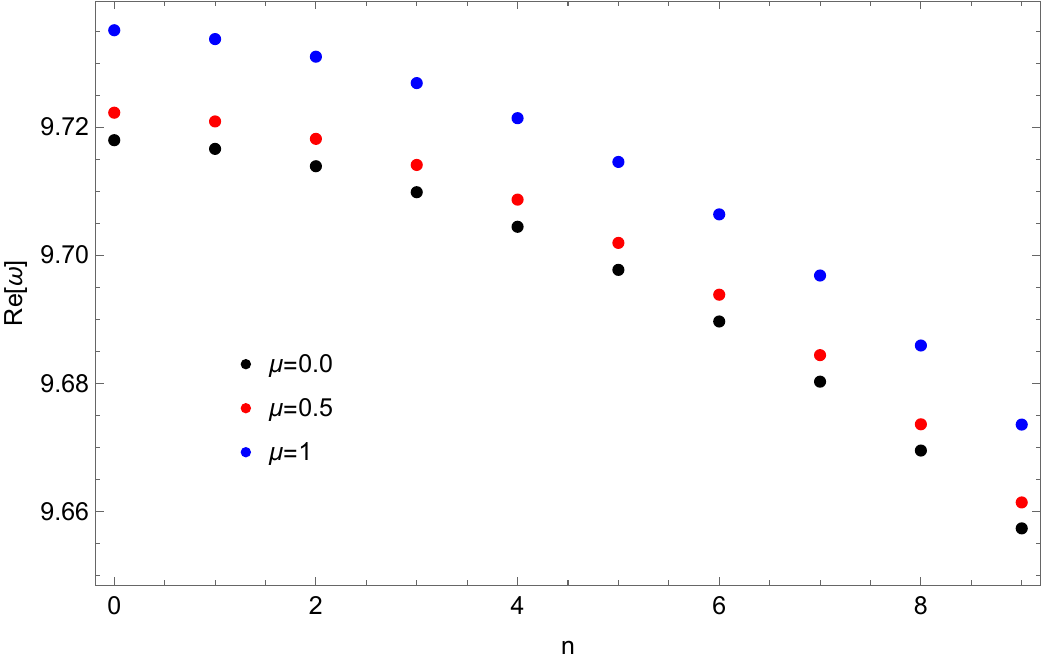}}
{\includegraphics[width=.49\textwidth]{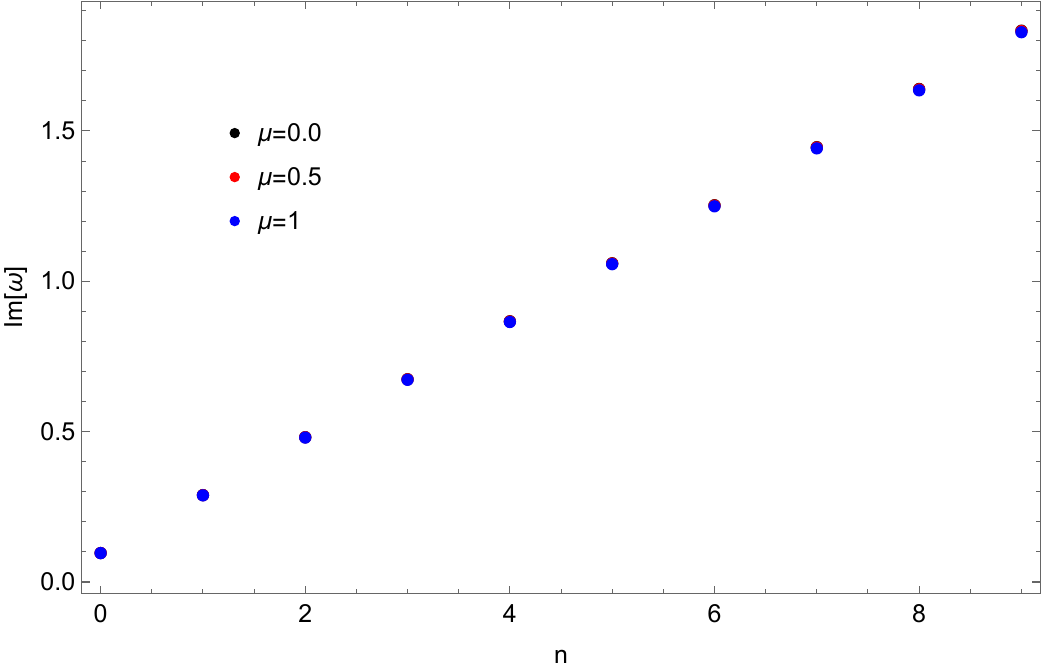}}
\caption{Real and imaginary parts of QNMs vs $n$ for different values of  $\mu$ when $\zeta = 50$, $\theta = 0.2$, and $\ell = 50$. }
\label{f4}
\end{figure}

Using this method implemented in a Mathematica package, we compute QNMs \(\omega\) for different values of \(\theta\), \(\zeta\), and \(\mu\), and analyze the behavior of the QNMs vs the overtone number \(n\) through Figs~(\ref{f2})-(\ref{f4}) schematically. The left panel of Fig.~(\ref{f2}) shows that for all values of the NC parameter $\theta$, $\text{Re}[\omega]$ decreases monotonically with increasing $n$. This behavior reflects the general trend that higher overtones oscillate at lower frequencies, consistent with the fact that black holes are not isolated systems and dissipate energy over time. Also it can be inferred when $\theta$ increases referring to the range (\ref{ran}), the values of real part of QNMs decrease across all overtones, indicating that noncommutativity suppresses the oscillation frequency of QNMs. This behavior is consistent with Fig.~(\ref{f1}c), where an increase in \(\theta\) results in a lower potential barrier, meaning that the wave requires less energy to overcome it. In addition, this suppression becomes more obvious at higher overtone numbers, suggesting that the NC effects are amplified for modes with larger $n$. Moreover, the plot reveals that even small changes in $\theta$ for instance, changing $\theta$ from $ 0.22$ to $0.23$ results in shifts in the spectrum, particularly for large $n$.

Due to energy loss in the black hole's dissipative system, the damping rate of the modes increases with the overtone number $n$, as shown by the monotonic growth of $-\text{Im}[\omega]$ in the right panel of Fig.~(\ref{f2}). While the real part is  affected by $\theta$, especially for larger $n$, the imaginary part remains almost insensitive to NC corrections within the considered parameter range.

Similarly, Fig.~(\ref{f3}) presents the real and imaginary parts of the QNMs for different values of the non-minimal coupling constant $\zeta$. As seen in the left panel, the effect of $\zeta$ is negligible for small overtone numbers, indicating that the coupling has little influence on the oscillation frequency at low $n$. However, as $n$ increases, $\text{Re}[\omega]$ decreases monotonically. The right panel shows that, much like $\theta$ in Fig.~(\ref{f2}), $\zeta$ does not have a strong effect on the imaginary part of the QNMs.

Fig.~(\ref{f4}) illustrates the effect of the scalar field mass $\mu$ on the QNM frequencies. As shown in the left panel, the dependence on $\mu$ already appears at the fundamental mode ($n = 0$), in contrast to the behavior observed for the parameters $\zeta$ and $\theta$, which only become significant at higher overtones. This indicates that $\mu$ plays a more direct role in shaping the spectral structure by modifying the effective potential. Meanwhile, as seen in the right panel, $-\text{Im}[\omega]$ exhibits no sensitivity to the mass, implying that the damping rate remains unaffected by $\mu$.

\section{Second scenario: Non-minimal derivative coupling to Einstein tensor}\label{sec3}

In this section, we explore a model in which the derivatives of the scalar field are non-minimally coupled to the Einstein tensor in the background of the NC-Sch black hole (see Eq.~(\ref{Sch})). The most general form of this type of coupling, involving the linear curvature tensors $R_{\mu\nu}$ and $R$, was proposed from an inflationary perspective in Ref.~\cite{Amendola:1993uh}. Although the scalar field equation derived from this Lagrangian is generally not of second order, it has been shown \cite{Sushkov:2009hk} that for a particular choice of couplings -- specifically, when the scalar field is coupled to the Einstein tensor -- the equation of motion reduces to a second-order differential equation. This coupling modifies the dynamics of scalar perturbations, leading to alterations in the QNM spectrum.

The corresponding non-minimally coupled scalar field theory is described by the following action \cite{Sushkov:2009hk}
\begin{equation}\label{ss}
S = -\frac{1}{2} \int d^4x\, \sqrt{-g} \left( g^{\mu\nu} \partial_\mu \phi \partial_\nu \phi + \mu^2 \phi^2 - \hat{\zeta} G^{\mu\nu} \partial_\mu \phi \partial_\nu \phi \right),
\end{equation}
where \(\hat\zeta\) is a dimensionful coupling constant, \(G^{\mu\nu}\) is the Einstein tensor and \(\mu\) denotes the mass of the scalar field. In the geometric units which \(M\) has the dimension of length, the dimensionful coupling \(\hat\zeta\) can be defined in terms of \(\zeta\) as \(\hat\zeta = M^2 \zeta\). However, in the rest of the numerical calculations when we set \(M=1\), both couplings are of the same values but not of the same dimensions.

Varying the action in Eq.~(\ref{ss}) with respect to the scalar field $\phi$ yields the equation of motion as follows:
\begin{equation}\label{Eom1}
\frac{1}{\sqrt{-g}} \partial_{\mu} \left[ \sqrt{-g} \left( g^{\mu \nu} - \zeta G^{\mu \nu} \right) \partial_{\nu} \phi \right] - \mu^{2} \phi = 0.
\end{equation}
Now to obtain the master equation, we first insert the following general ansatz in Eq.~(\ref{Eom1}) 
\begin{equation} \label{ansatz1}
\phi(t,r,\vartheta,\varphi) = B(r) Y_{\ell m}(\vartheta,\varphi)\, e^{-i \omega t},
\end{equation}
then up to an overall factor $2r^3 f(r) $ the radial differential equation becomes 
\begin{equation} \label{req}
\frac{1}{r^2} \frac{d}{dr} \left[ r^2 f(r) \bigl(1 + \zeta K(r) \bigr) B'(r) \right] +\frac{\omega^2\bigl(1 + \zeta K(r) \bigr)}{f(r)} B(r)-\left( \mu^2 + \frac{\ell(\ell+1)}{r^2} \bigl(1 - \zeta L(r)\bigr) \right)B(r) = 0,
\end{equation}
which is similar to Eq.~(\ref{reqs}) and 
\begin{equation} \label{kl}
K(r) = \frac{1 - f(r)}{r^2} - \frac{f'(r)}{r}, \qquad L(r) = K(r) - \frac{1}{2} R.
\end{equation}
Here, $R$ is the Ricci scalar of the background geometry and the prime is the derivative with respect to the $r$ coordinate. This equation can be recast for convenience in the following form 
\begin{equation} \label{E1}
f(r)^2 B''(r) + \Sigma(r) B'(r) +\left(\omega^2+ \Xi(r)\right) B(r) = 0,
\end{equation} 
where
\begin{align}
\Sigma(r) &= f(r)^2 \left( \frac{\zeta K'(r)}{1+\zeta K(r)} + \frac{f'(r)}{f(r)} + \frac{2}{r} \right), \label{H1} \\
\Xi(r) &= \frac{-f(r)}{1+\zeta K(r)} \left( \frac{\ell(\ell+1)}{r^2}\bigl(1 - \zeta L(r)\bigr) + \mu^2 \right). \label{H3}
\end{align}

Following the same procedure in the first scenario, by using the field redefinition
\begin{equation} \label{Xi}
B(r) = \frac{1}{r \left(1 + \zeta K(r) \right)^{1/2}} \psi(r),
\end{equation}
 and the tortoise coordinate $r^{*}$ given by Eq.~(\ref{tc}), we can write Eq.~(\ref{E1}) in the form of a Schr\"{o}dinger-like equation as follows:
\begin{equation} \label{SE}
\frac{d^2 \psi(r)}{d{r^{*}}^2} + \left( \omega^2 - V(r) \right) \psi(r) = 0,
\end{equation}
where the effective potential $V(r)$ takes the form
\begin{equation} \label{VT}
V(r) = \frac{f(r) f'(r)}{r} + f(r)^2 H(r) - \Xi(r),
\end{equation}
and the function $H(r)$ is defined by
\begin{equation}
H(r) = \frac{\zeta}{1+\zeta K(r)} \left( \frac{K''(r)}{2} + \frac{K'(r) f'(r)}{2f(r)} + \frac{K'(r)}{r} \right) - \frac{1}{4} \left( \frac{\zeta K'(r)}{1+\zeta K(r)} \right)^2. \label{H2}
\end{equation}

\begin{figure}[H]
\centering
\subfigure[$\mu=0.5$, $\zeta=40$, $\ell=40$, $\theta=0.12$]
{\includegraphics[width=.32\textwidth]{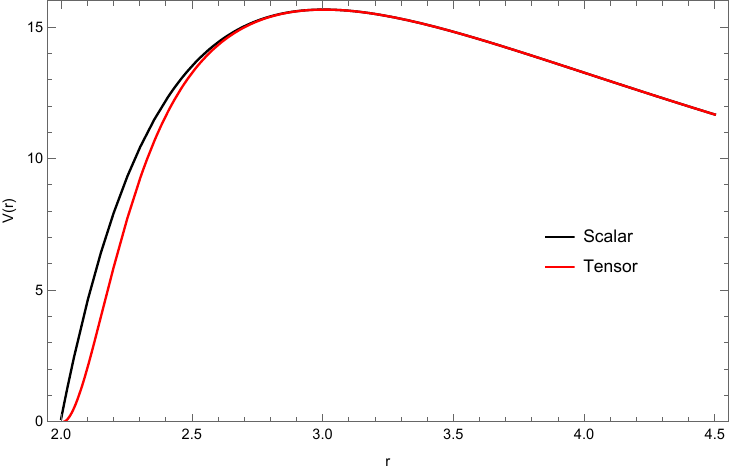}}
\subfigure[$\mu=0.5$, $\zeta=40$, $\ell=40$, $\theta=0.15$]
{\includegraphics[width=.32\textwidth]{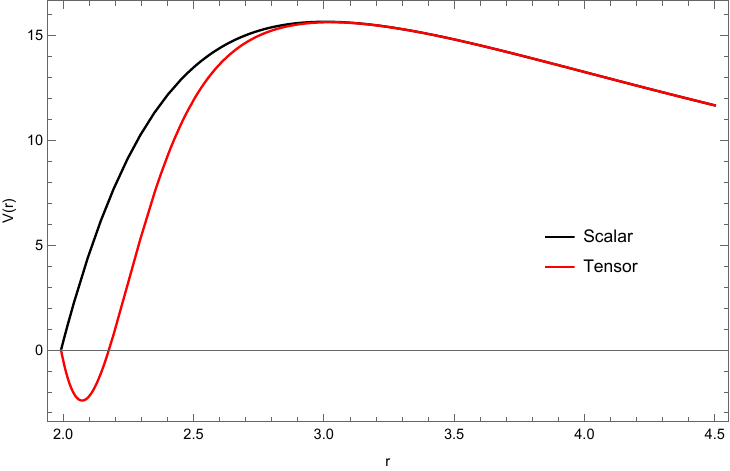}}
\subfigure[$\mu=0.5$, $\zeta=150$, $\ell=40$, $\theta=0.12$]
{\includegraphics[width=.32\textwidth]{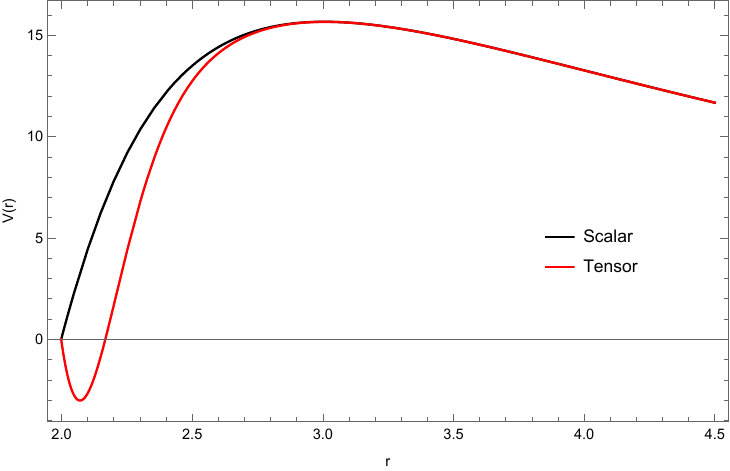}}
\caption{Comparison between the effective potentials of two models. }
\label{f5}
\end{figure}

To compare the effective potentials \( V(r) \) for the two scenarios, we present Fig.~(\ref{f5}). The main differences appear in the near-horizon region, specifically within the interval \( 2M < r < 3M \), while for \( r > 3M \) the two potentials converge and the discrepancy becomes negligible. As shown in the figure, the difference in the potential profiles diminishes for small values of \( \zeta \) and \( \theta \), indicating that the deviation is governed by the strength of the coupling and the NC parameter. Thus, the distinction between the two models is most visible in regions close to the horizon and for specific ranges of the parameters.

 \subsection{QNMs in the Tensor model}

\begin{figure}[H]
\centering
\subfigure[]
{\includegraphics[width=.47\textwidth]{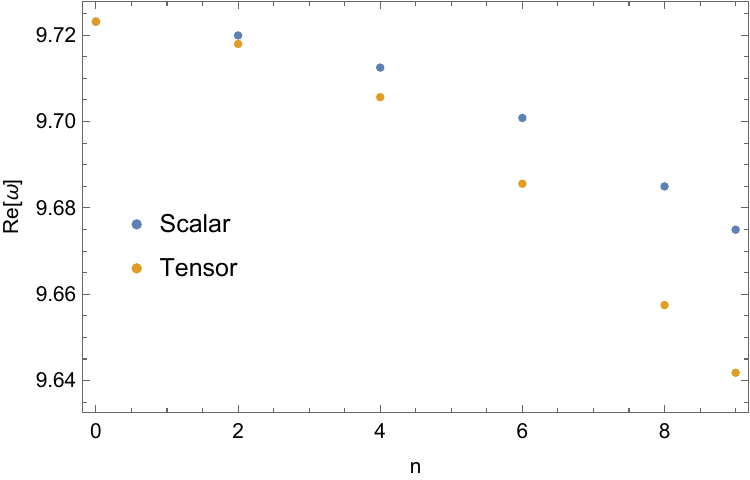}}
\subfigure[]
{\includegraphics[width=.47\textwidth]{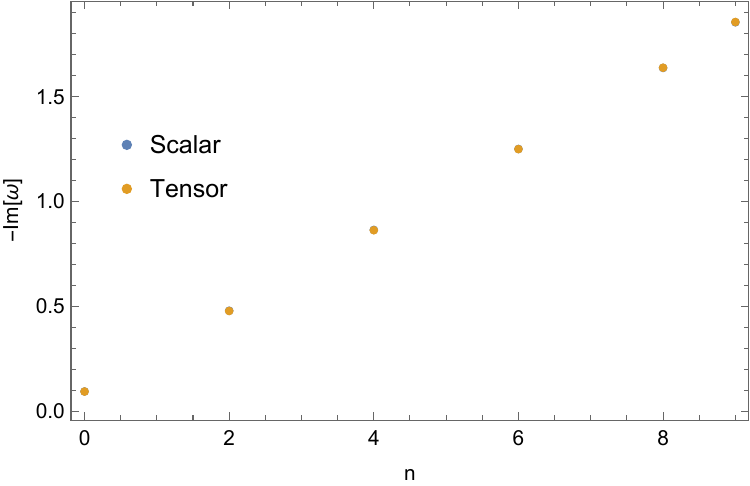}}
\caption{Comparison between the QNMs of two scenarios for $\mu=0.5$, $\zeta=20$, $\ell=50$ and $\theta=0.12$. }
\label{f6}
\end{figure}

Following the sixth-order WKB approximation alluded to in the previous section, we compute the QNM frequencies associated with scalar perturbations in the second scenario. Our calculations reveal that the general behavior is similar to that of the first scenario, presented in Figs.~(\ref{f2}--\ref{f4}). However, in order to highlight the differences between the two models, we present Fig.~(\ref{f6}), which provides a direct comparison of the QNM frequencies for appropriate parameter values.

In Fig.~(\ref{f6}), we observe that for low overtones — particularly for the fundamental mode ($n = 0$) — both models yield nearly identical results for both $\text{Re}[\omega]$ and $\text{Im}[\omega]$. This indicates that in the low-frequency regime, the dynamical response of the system is insensitive to the specific form of the coupling between the scalar field and the gravitational background. However, as the overtone number increases, a clear distinction between the two models emerges.

\subsection{Convergence of WKB approximation}

To assess the numerical reliability of the QNM frequencies obtained within the WKB approximation, the convergence of the method is examined in this subsection. As a representative case, the tensor-coupled model is
considered. The corresponding frequencies and estimated
errors are presented in Tab.~\ref{qnm}.
One can see that the real and imaginary parts of the frequencies remain stable as the order of the approximation is increased from three to six. 

\begin{table}[htbp]
\centering
\caption{Estimated errors for different WKB orders at $n=9$ when $\theta=0.2$, $\zeta=50$, $\ell=50$, and $\mu=0.5$.}
\label{qnm}
\begin{tabular}{ccc}
\hline
WKB order & QNM & Estimated error \\
\hline
3 & $9.66111 - 1.83067\,i$ & $0.00486771$ \\
4 & $9.66232 - 1.83044\,i$ & $0.000991559$ \\
5 & $9.66261 - 1.83197\,i$ & $0.000991847$ \\
6 & $9.66140 - 1.83220\,i$ & $0.000704197$ \\
\hline
\end{tabular}
\end{table}

We have also investigated the estimated errors for the sixth-order of WKB method via the overtone numbers in Tab.~\ref{qnm2}. For the considered parameter space, the real parts of QNM decrease gradually by increasing the overtone number, while the magnitude of the imaginary parts increase as well, indicating strong damping in the higher overtones. The estimated error
likewise increases with $n$, which is consistent with the increasing difficulty
of obtaining accurate WKB results for highly damped overtones.

\begin{table}[htbp]
\centering
\caption{Estimated errors of the sixth-order WKB approximation for different overtone numbers when $\theta=0.2$, $\zeta=50$, $\ell=50$, and $\mu=0.5$.}
\label{qnm2}
\begin{tabular}{ccc}
\hline
$n$ & QNM & Estimated error \\
\hline
0 & $9.722262023208 - 0.09624805009419\,i$ & $5.84086\times10^{-10}$ \\
1 & $9.7208998218 - 0.2887513237\,i$ & $4.87659\times10^{-8}$ \\
2 & $9.71817907 - 0.48127714\,i$ & $4.88126\times10^{-7}$ \\
3 & $9.71410625 - 0.67384405\,i$ & $2.59713\times10^{-6}$ \\
4 & $9.70868893 - 0.86647604\,i$ & $9.81256\times10^{-6}$ \\
5 & $9.701933 - 1.059205\,i$ & $2.96397\times10^{-5}$ \\
6 & $9.693841 - 1.252074\,i$ & $7.63432\times10^{-5}$ \\
7 & $9.684410 - 1.445142\,i$ & $1.74528\times10^{-4}$ \\
8 & $9.673605 - 1.638484\,i$ & $3.63738\times10^{-4}$ \\
9 & $9.661403 - 1.832202\,i$ & $7.04197\times10^{-4}$ \\
\hline
\end{tabular}
\end{table}

In order to improve the convergence of the WKB expansion, we use the Pad{\'e} approximations suggested first by J. Matyjasek and M. Opala \cite{Matyjasek:2017psv}. The resulting frequencies at successive WKB
orders for $n=9$ are given in Tab.~\ref{qnm3}. In comparison with the standard WKB results in Tab.~\ref{qnm}, the Pad{\'e}-resummed frequencies exhibit smaller estimated errors at each order. However, the convergence of the Pad{\'e}-resummed sequence is not strictly monotonic which can be raised from the rational resummation of a truncated asymptotic WKB series and does not necessarily indicate a loss of numerical stability.

\begin{table}[htbp]
\centering
\caption{Estimated errors for different WKB orders using the Pad{\'e} approximation for $n=9$, $\theta=0.2$, $\zeta=50$, $\ell=50$, and $\mu=0.5$.}
\label{qnm3}
\begin{tabular}{ccc}
\hline
WKB order & QNM & Estimated error \\
\hline
3 & $9.66131 - 1.83064\,i$ & $0.00020143$ \\
4 & $9.66233 - 1.83053\,i$ & $0.0000670908$ \\
5 & $9.66189 - 1.83151\,i$ & $0.000494208$ \\
6 & $9.66159 - 1.83162\,i$ & $0.000199113$ \\
\hline
\end{tabular}
\end{table}

Finally, the dependence of QNM and estimated errors to the overtone numbers $n=0,\ldots,9$ for the sixth-order WKB method accompanied with Pad{\'e}-resummed are listed in Table~\ref{qnm4}. The Pad{\'e} resummation substantially reduces the estimated errors compared with the corresponding sixth-order WKB results in Table~\ref{qnm2}, with the improvement being particularly pronounced for the lower overtones. Even for the higher overtones, for which the WKB approximation becomes less accurate, the Pad{\'e}-resummed results provide a smaller estimated error for the parameter set
considered here.

\begin{table}[htbp]
\centering
\caption{Estimated errors of the sixth-order WKB-Pad{\'e} approximation for different overtone numbers when
$\theta=0.2$, $\zeta=50$, $\ell=50$, and $\mu=0.5$.}
\label{qnm4}
\begin{tabular}{ccc}
\hline
$n$ & QNM & Estimated error \\
\hline
0 & $9.7222620232089 - 0.0962480500981\,i$ & $8.50672\times10^{-12}$ \\
1 & $9.72089984420 - 0.2887512830528\,i$ & $1.07008\times10^{-9}$ \\
2 & $9.71817919259 - 0.4812768199900\,i$ & $4.56173\times10^{-9}$ \\
3 & $9.714106869 - 0.6738422579\,i$ & $3.65343\times10^{-8}$ \\
4 & $9.70869035 - 0.866470304\,i$ & $2.06103\times10^{-6}$ \\
5 & $9.7019382 - 1.05918617\,i$ & $6.65644\times10^{-6}$ \\
6 & $9.6938557 - 1.2520215\,i$ & $1.83691\times10^{-5}$ \\
7 & $9.6844432 - 1.4450120\,i$ & $4.47080\times10^{-5}$ \\
8 & $9.673694 - 1.63819682\,i$ & $9.83313\times10^{-5}$ \\
9 & $9.661591 - 1.83161695\,i$ & $1.99113\times10^{-4}$ \\
\hline
\end{tabular}
\end{table}

\section{Ringdown profiles}\label{sec4}

An alternative approach to investigate the influence of curvature couplings on
the dynamical evolution of a scalar field in the NC-Sch black hole spacetime is
to study the time evolution of perturbation profiles using the standard  time-domain integration method proposed by Gundlach--Price--Pullin
\cite{Gundlach:1993tn}. This technique has been widely
employed in studies of QNMs and time-domain profiles of scalar fields
\cite{Abdalla:2006vb,Chirenti:2007mk,Toshmatov:2017bpx}. We discrete the wave equations (\ref{sceq}) and (\ref{SE}) for each scenario using a finite-difference method. During the integration, we extract the field values at constant $r^{*}$ and evolve the field to larger $t$, subsequently plotting the ringdown profiles. For numerical convenience, we adopt the light-cone (null) coordinates defined by
\begin{equation}
u = t - r^{*}, \qquad v = t + r^{*},
\end{equation}
so that, with appropriate initial conditions specified on the null grids, the
corresponding wave equation can be integrated to obtain the time-domain
profiles. In these coordinates, the wave equation takes the form
\begin{equation}
\left(4 \frac{\partial^2}{\partial u \partial v} + V_s(u, v)\right) \Psi(u, v) = 0,
\end{equation}
where $V_s(u, v)$ denotes the effective potential, which depends on the
parameters of the theory and the background geometry. By evolving the
discretised wave function step by step on the $(u, v)$ grid, subject to initial
conditions specified along the two null surfaces $u = u_0$ and $v = v_0$, we
obtain the time-domain evolution of the scalar perturbations.

Following \cite{Gundlach:1993tn}, we discretise the wave equation on a uniform grid
$u_i=i\,\Delta$, $v_j=j\,\Delta$ in null coordinates, with $r_*=(v-u)/2$ and
$t=(u+v)/2$. The second-order accurate stencil reads
\begin{equation}\label{eq:gpp_stencil}
\psi_{i+1,j}=\psi_{i+1,j-1}+\psi_{i,j}-\psi_{i,j-1}
-\frac{\Delta^2}{8}\,V_{i+1/2,j-1/2}\bigl(\psi_{i+1,j-1}+\psi_{i,j}\bigr),
\end{equation}
where the potential is evaluated at the centre of the cell, i.e.\ at
$r_*=(j-i-1)\Delta/2$. The grid spacing used for the displayed waveforms is
$\Delta=0.3$ and the evolution extends up to $t_{\max}=600$. To verify the
accuracy of the scheme, we repeated the evolution with
$\Delta=0.3,\,0.15,\,0.075$ on the same physical domain.

Since the basic features of the field decay are independent of the initial
conditions, we choose a Gaussian wave packet centered at $v=v_c$ with width
$\sigma$ as the initial disturbance,
\begin{equation}\label{ip}
\psi(u=u_0,v)=a\exp\left[-\frac{(v-v_c)^2}{2\sigma^2}\right],
\end{equation}
set on the null surface $u=0$ with $a=1$, $v_c=15$ and $\sigma=3$, in accordance
with the choices used in the literature. On the second null surface $v=0$ we set
$\psi(u,0)=0$. Since the scheme is characteristic, no boundary condition is
needed at large $r_*$; the field is extracted at $r_*=0$  and  at $r_*=24.9$ .

\begin{figure}[htbp]
\centering
\includegraphics[width=0.5\textwidth]{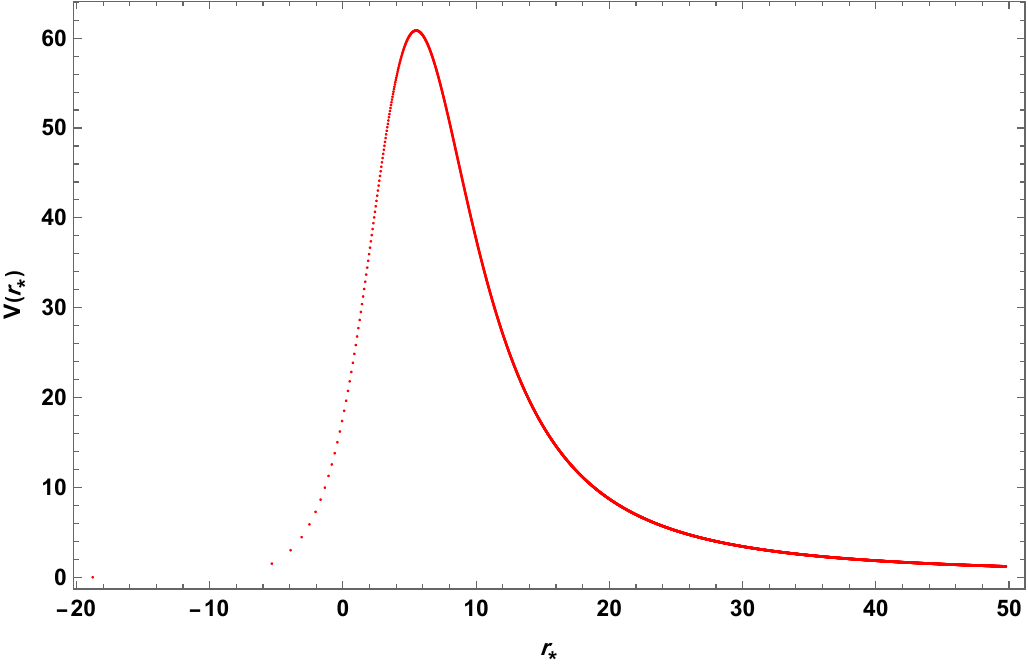}
\caption{Effective potential of scalar model vs. $r_{*}$ when $\ell=40$, $\theta=0.12$,
$\mu=0.5$, and $\zeta=40$.}
\label{ep}
\end{figure}

To evaluate $V(r_*)$ on the grid we need $r$ as a function of $r_*$. We fix the
additive constant of the tortoise coordinate by
\begin{equation}\label{eq:tortoise_integral}
r_*=\int_{r_0}^{r}\frac{dr'}{f(r')},\qquad r_0=2.1,
\end{equation}
and obtain $r(r_*)$ by integrating $dr/dr_*=f(r)$ with an adaptive 8th-order
Runge--Kutta method (relative and absolute tolerances $10^{-12}$ and $10^{-14}$)
exactly at the grid nodes $r_*=k\Delta/2$, so that neither interpolation nor a
horizon offset is needed. The potential is then evaluated analytically in double
precision, and is set to zero for $r_*<-50$, where it is exponentially small
($V\sim e^{r_*/2.4}\lesssim10^{-9}$). The
resulting potential for a representative set of parameters is shown in
Fig.~(\ref{ep}).

\begin{figure}[H]
\centering
\subfigure[Ringdown profiles]
{\includegraphics[width=.52\textwidth]{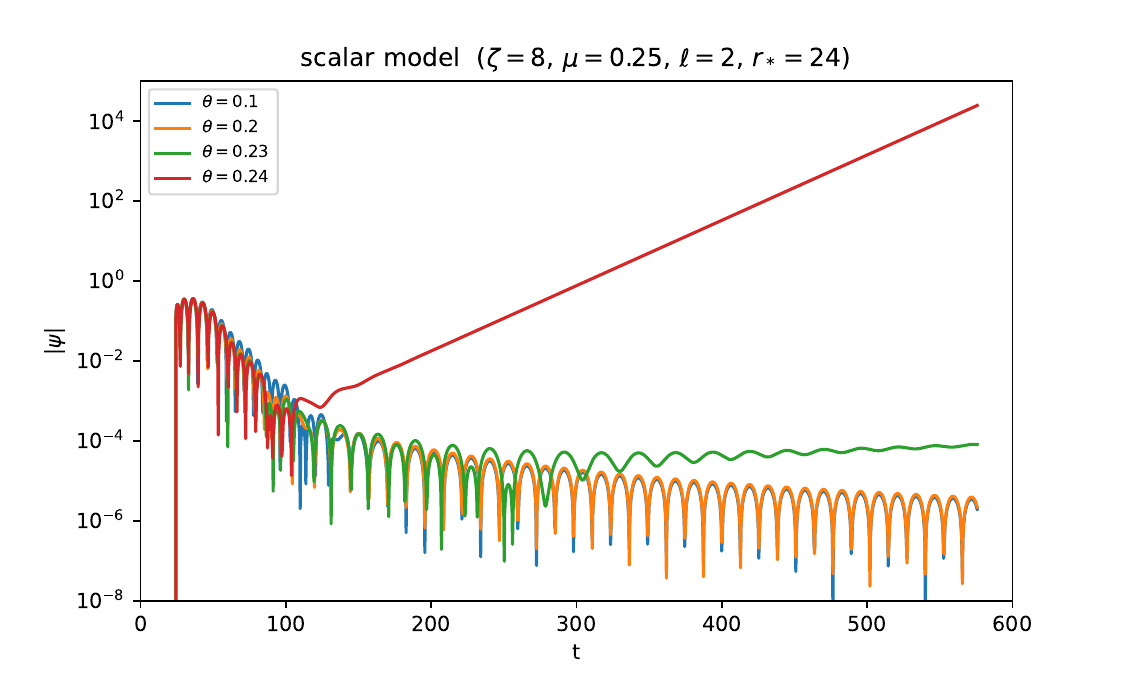}}
\subfigure[Effective potential]
{\includegraphics[width=.44\textwidth]{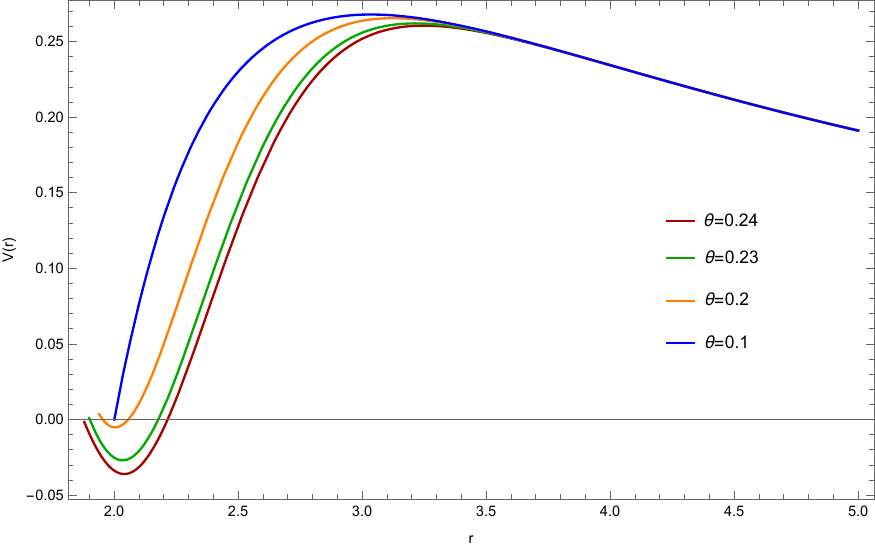}}
\caption{\(|\psi| \) and \(V( r) \) for different values of \( \theta \) when \( \ell = 2 \), \( \zeta = 8 \), and \( \mu = 0.25 \). }
\label{f7}
\end{figure}

In the left panel of Fig.~(\ref{f7}), we plot the ringdown profiles for different values of the NC parameter $\theta$ in the scalar model. For small values of $\theta$, the waveform exhibits the standard features of quasi-normal ringing: an initial burst, followed by a damped oscillatory phase, and then a transition to a late-time tail. As $\theta$ increases up to values denoted by (\ref{ran}), the damping rate diminishes and beyond a certain value of $\theta$ in particular, for $\theta = 0.23$, the logarithmic amplitude grows exponentially at late times, signaling a transition to an unstable regime. This behavior indicates that there exists a critical value beyond which the NC deformation triggers instability in the scalar evolution.

To further illustrate the onset of instability at large $\theta$, we present the effective potential $V(r)$ for different $\theta$ in the right panel of Fig.~(\ref{f7}). As is evident from the plot, increasing $\theta$ leads to the development of a progressively deeper negative well in the potential. For sufficiently large $\theta$, this well becomes deep enough to support bound states with negative energy, which signals the presence of an unstable mode. It is worth mentioning that the emergence of instability here is originated from the late-time behaviour of the ringdown profiles not especially the appearance of the well potential. It has been shown in Refs.~\cite{Cardoso:2009cnd,Cardoso:2013fwa} that for the stability, it can be sufficient to have $\int \frac{V(r)}{f(r)} dr>0$ but this is not a necessary condition because our values does not yield negative integral. However, it should not be interpreted that the positivity of this integral ignores negative potential values, which could correspond to bound states with negative eigenvalues. In fact, the perturbative mode goes to instability before this integral become negative because the non-negativity of the potential is a sufficient condition not the positivity of its integral.

\begin{figure}[H]
\centering
\subfigure[Ringdown profiles]
{\includegraphics[width=.52\textwidth]{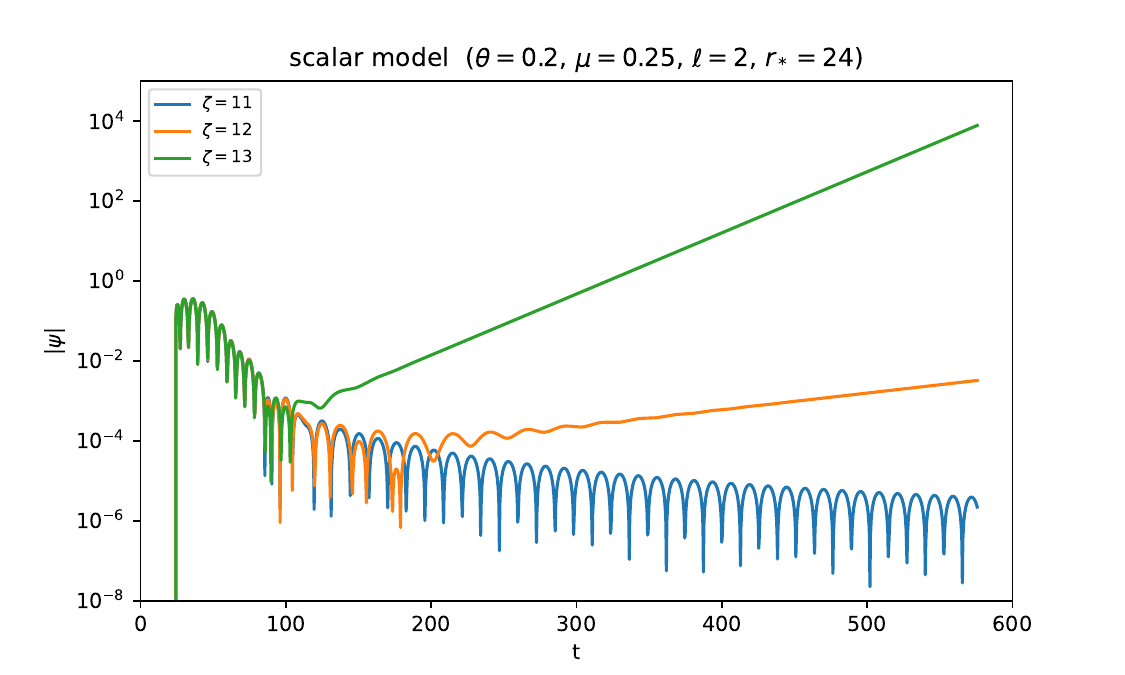}}
\subfigure[Effective potential]
{\includegraphics[width=.44\textwidth]{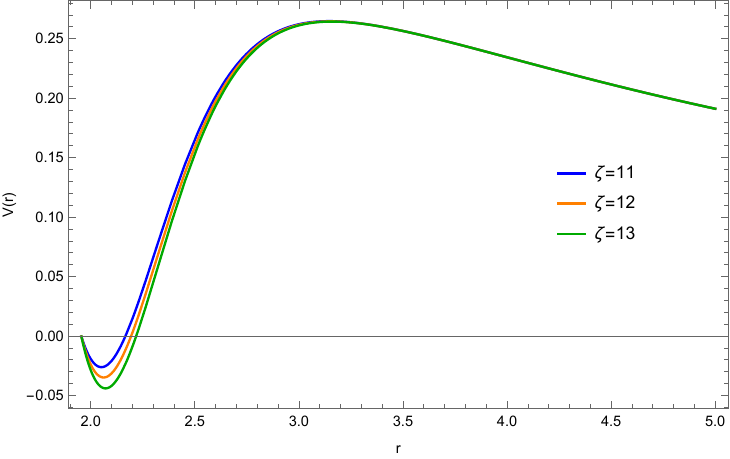}}
\caption{\(|\psi| \) and \(V( r) \) for different values of \( \zeta \) when \( \theta = 0.2 \), \( \mu = 0.25 \) and \( \ell = 2 \). }
\label{f8}
\end{figure}

A similar instability emerges for large values of the coupling constant $\zeta$. As shown in Fig.~(\ref{f8}a), for small and moderate values of $\zeta$, the ringdown profile exhibits the typical exponentially decaying oscillations as characteristic of stable QNMs. However, as $\zeta$ increases, the damping rate diminishes, and eventually the amplitude begins to grow with time -- a clear signature of instability. To gain insight into this transition, we examine the effective potential $V(r)$ in the right panel. One observes that increasing $\zeta$ progressively deepens a negative-valued well near the horizon.

Fig.~(\ref{f9}) shows that the ringdown profiles of a scalar field with mass $\mu=0$, $0.25$ and $0.5$ for $\theta=0.2$ and $\ell=2$ in the scalar-coupling model, extracted at $r_*=24$. In the left panel for $\zeta=5$, all three profiles are stable, however after ringing, the massless field settles into a non-oscillatory power-law tail, $|\psi|\propto t^{-(2\ell+3)}=t^{-7}$ (the black dashed line on the blue curve), whereas for $\mu\neq 0$ the late-time signal is an oscillatory tail with period $2\pi/\mu$ that decays more slowly; the larger the mass, the slower the decay \cite{Konoplya:2018qov,Konoplya:2022hll}. 

In the right panel, we have plotted its behavior again for larger coupling constant $\zeta=12$. After the initial ringing, the profiles with $\mu=0$ and $\mu=0.25$ enter an exponentially growing phase, signalling the onset of the instability, while the one with mass $\mu=0.5$ remains stable. That is its late-time signal is an oscillating tail of nearly constant amplitude with no exponential growth. This behaviour follows from the structure of the effective potential. The curvature coupling $\zeta R$ produces a negative well localised near the horizon, whereas the mass contributes $\mu^2 f(r)$, which is positive and tends to $\mu^2$ at large $r$. \footnote{It should be noted that in secs.~ \ref{sec2} and \ref{sec3}, we found the QNM spectra from WKB method which yields satisfactory and careful results only when restricted by lower overtones $\ell>n$, but in this section we investigate the instability regime which is not necessary to implement this inequality. However, one can also find the QNM from ringdown profiles even when $\ell > n$ but this is not our goal in this paper.} 

Though the ringdown profiles presented above correspond to the scalar coupled model, we have verified that the qualitative effects of changing the parameters $\theta$, $\zeta$, and $\mu$ on the ringdown evolution are the same for the tensor coupled model. In contrast, the dependence on the multipolar number $\ell$ exhibits a completely different behavior in the two scenarios:

 \begin{figure}[H]
\centering
\subfigure[  \( \zeta = 5 \)]
{\includegraphics[width=.49\textwidth]{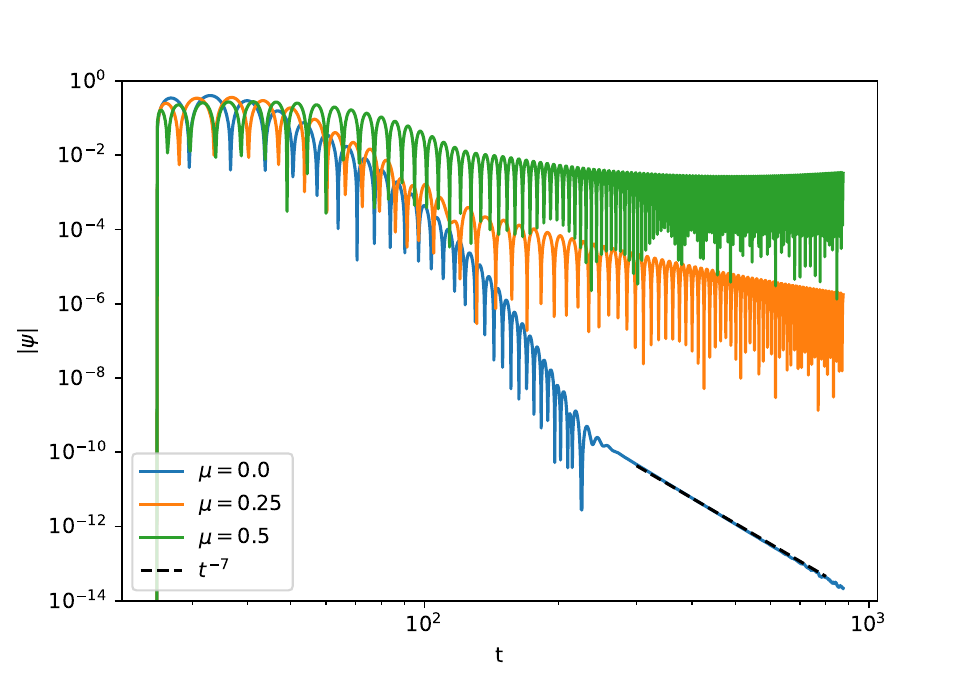}}
\subfigure[\( \zeta = 12 \) ]
{\includegraphics[width=.49\textwidth]{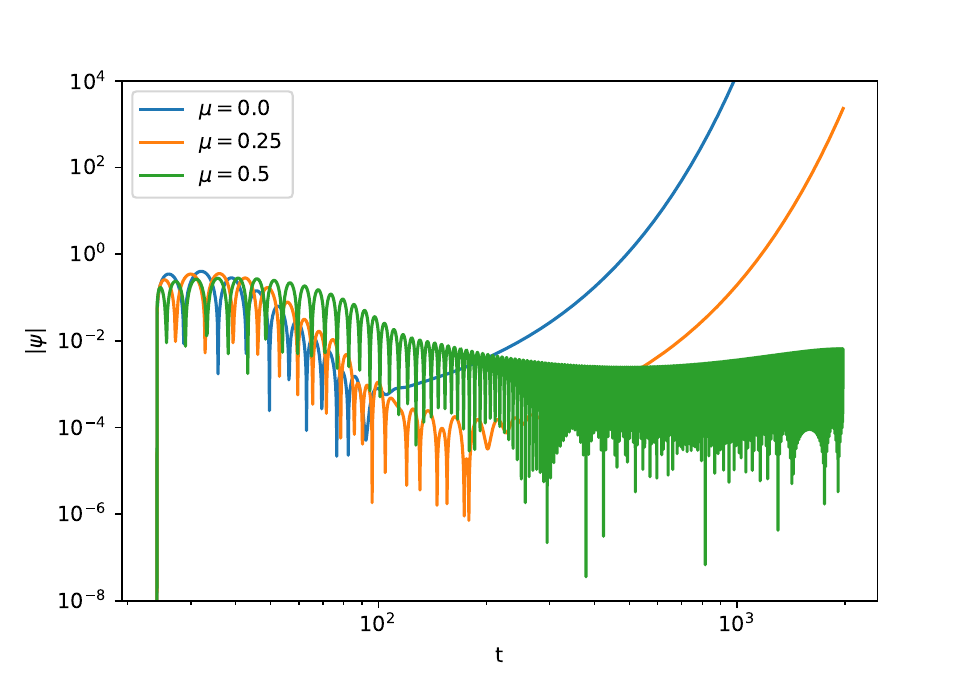}}
\caption{Ringdown profiles for different values of $\mu$ when \( \theta = 0.2 \)  and \( \ell = 2 \). }
\label{f9}
\end{figure}

\begin{figure}[H]
\centering
{\includegraphics[width=.55\textwidth]{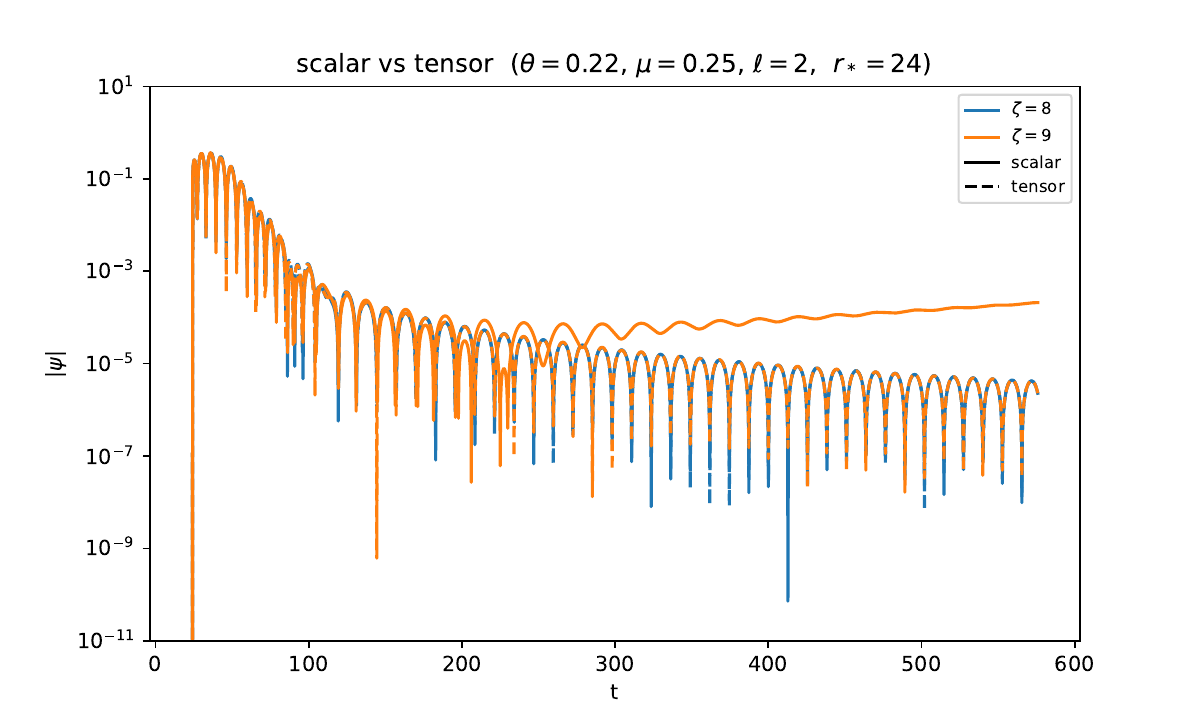}}
\caption{Ringdown profiles for two scenarios at  \( \ell = 2 \) }
\label{f13}
\end{figure}

\begin{figure}[H]
\centering
{\includegraphics[width=.55\textwidth]{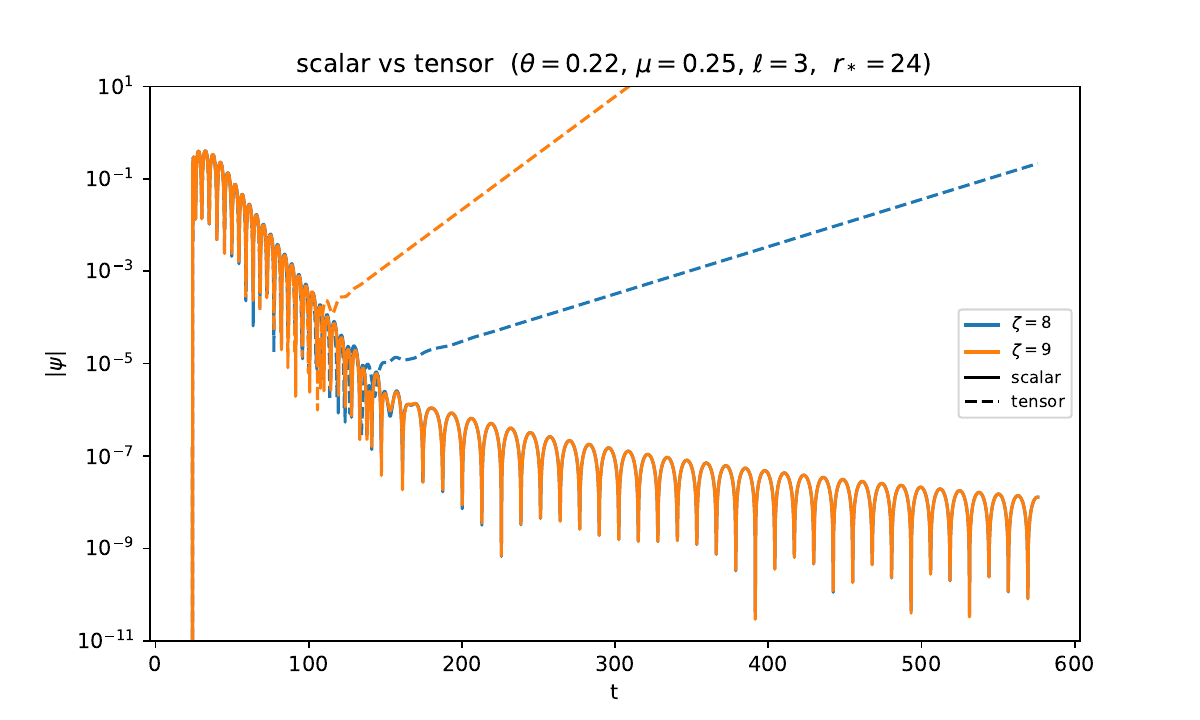}}
\caption{Ringdown profiles for two models at \(\ell=3\) }
\label{f14}
\end{figure}

\begin{itemize}
\item According to Fig.~(\ref{f10}) for the scalar model, when $\ell$ is small the effective potential contains a sufficiently deep well to support quasi‑bound states. These states lead to a prolonged ringdown and an overall unstable behavior, characterized by slow decay or even growth of perturbations. As $\ell$ increases, two competing effects come into play: (i) the height of the potential barrier increases, suppressing transmission to infinity, and (ii) the depth of the potential well decreases. The net effect is a reduction in the lifetime of any trapped mode. For sufficiently large $\ell$, the well disappears entirely, and the potential becomes purely barrier-like. In this regime, no bound states can form, and the system is linearly stable. The absence of late‑time tails for large $\ell$ further confirms that the power-law decay -- reported in the literature -- is not a generic feature of this coupling.

\begin{figure}[H]
\centering
\subfigure[Ringdown profiles]
{\includegraphics[width=.52\textwidth]{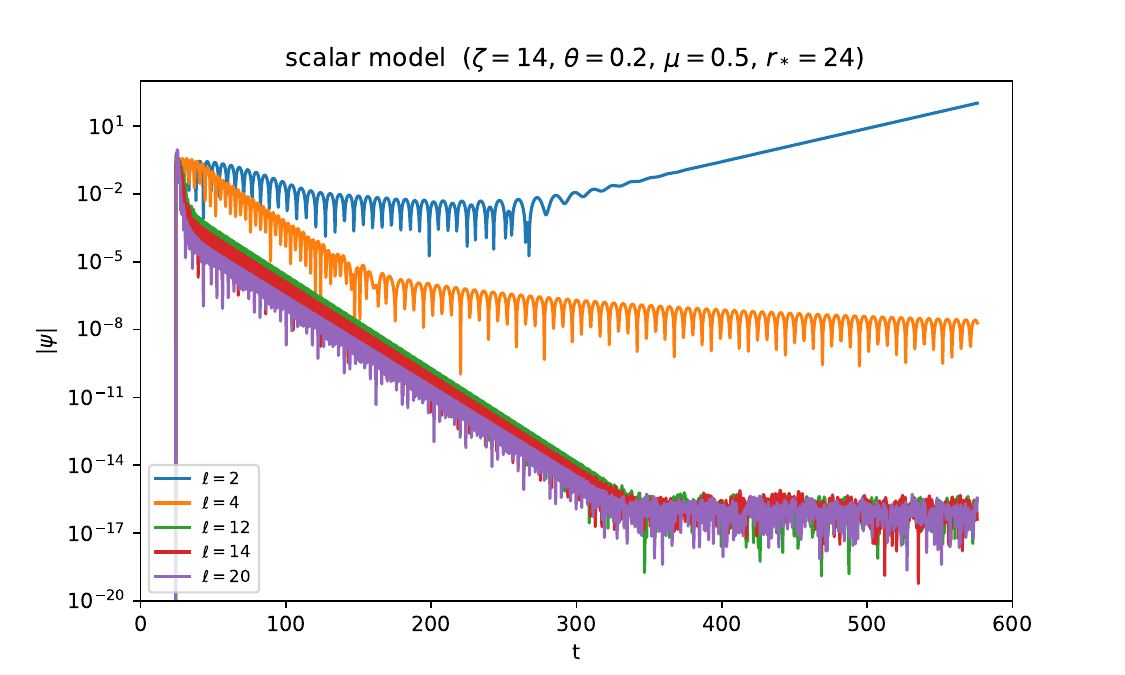}}
\subfigure[Effective potential]
{\includegraphics[width=.44\textwidth]{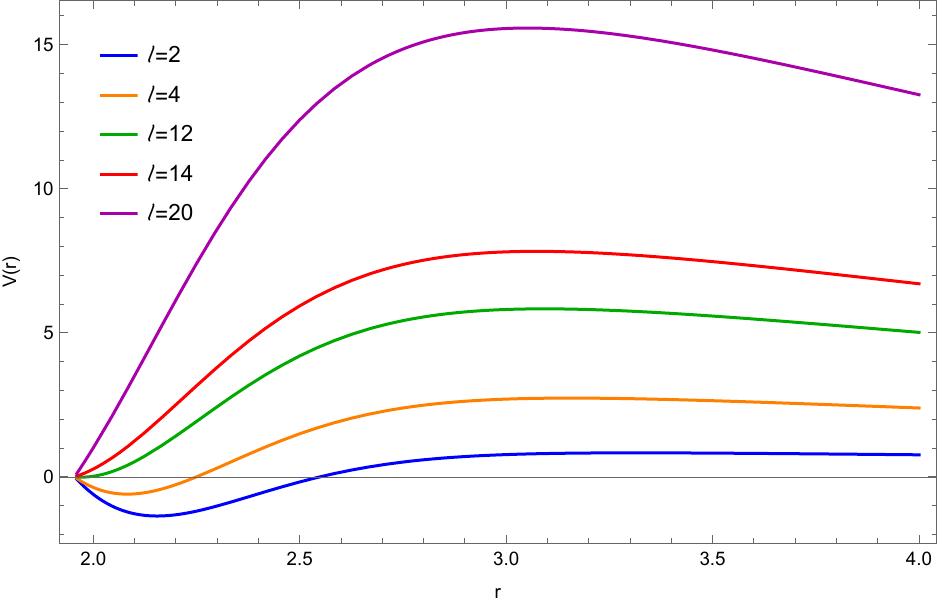}}
\caption{Ringdown profiles of scalar model for different values of $\ell$. }
\label{f10}
\end{figure}

\item On the other hand, for the tensor model shown in Fig.~(\ref{f11} b), although the barrier height again increases with $\ell$, the potential well deepens monotonically as $\ell$ grows. This occurs because the derivative interaction introduces an effective attractive term proportional to $\ell(\ell+1)$. Hence, higher multipolars experience a stronger binding force. The deepening well enhances the formation of long‑lived bound states, driving the system toward instability as $\ell$ increases.

  \begin{figure}[H]
\centering
\subfigure[Ringdown profiles]
{\includegraphics[width=.52\textwidth]{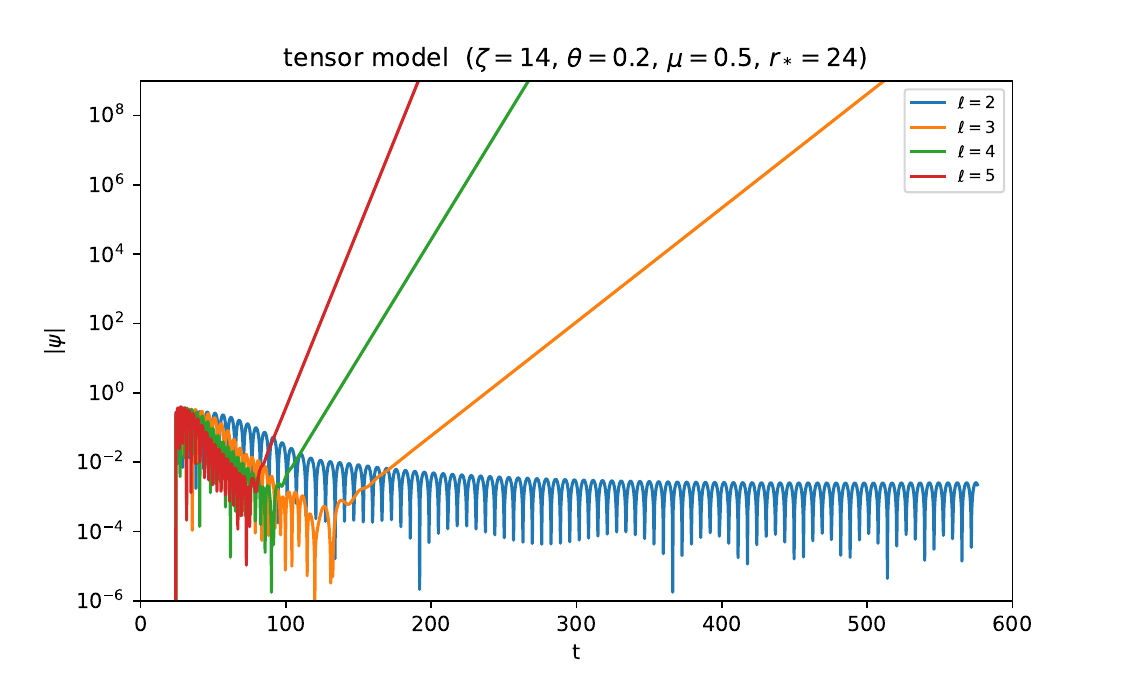}}
\subfigure[Effective potential]
{\includegraphics[width=.44\textwidth]{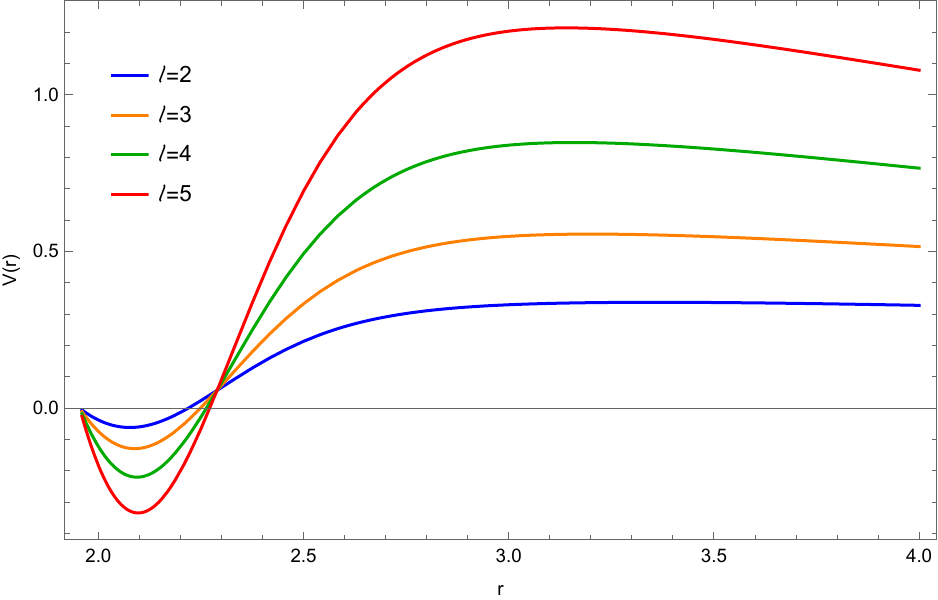}}
\caption{Ringdown profiles of Tensor model for different values of $\ell$  }
\label{f11}
\end{figure}
\end{itemize}
As a main motivation in this paper, here we provide a brief comparison of two non-minimal coupling scenarios for different values of parameter space. As can be seen from Figs.~(\ref{f13}) and (\ref{f14}), both models behave almost identically at $\zeta=8$. However, as $\zeta$ is increased from $8$ to $9$, two different outcomes emerge. At $\zeta=9$, as shown in Fig.~(\ref{f13}), for $\ell=2$ the tensor model is more stable than the scalar one. When $\ell$ is increased to $3$ in Fig.~(\ref{f14}), however, the situation is completely reversed and the scalar model turns out to be the more stable one, which confirms the earlier results discussed above.

\begin{figure}[H]
\centering
{\includegraphics[width=.49\textwidth]{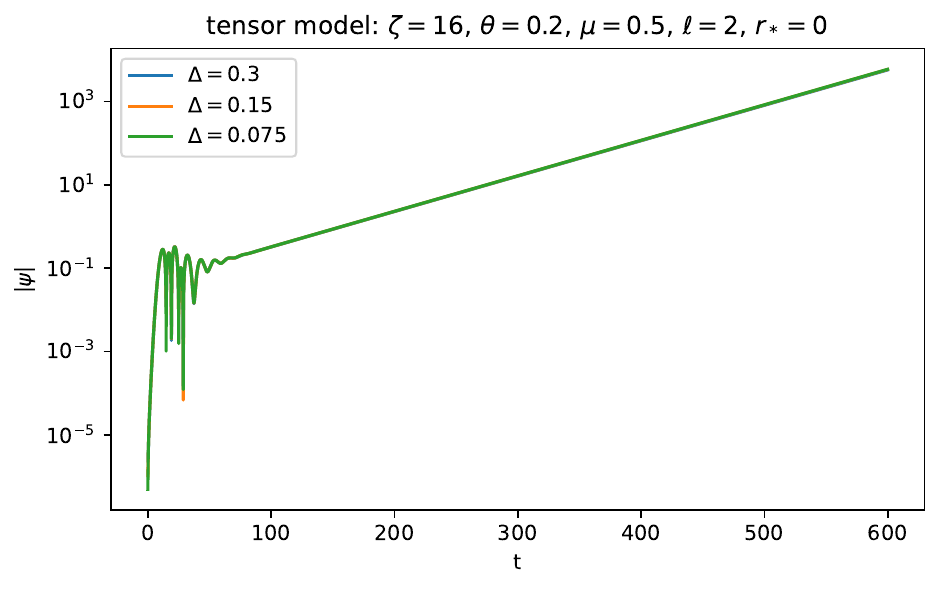}}
{\includegraphics[width=.49\textwidth]{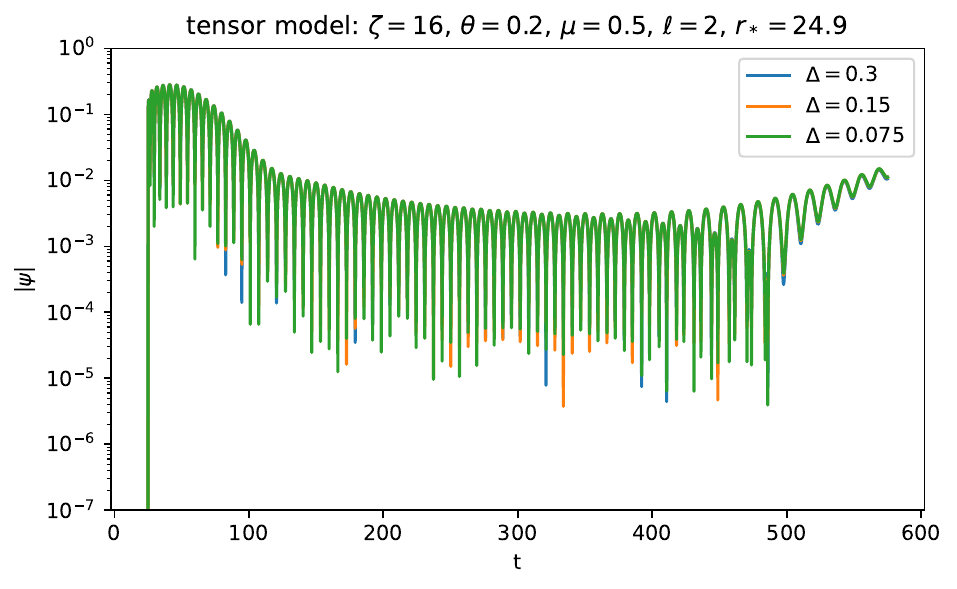}}
\caption{Ringdown profiles of the tensor model for different grid resolutions.  }
\label{conv}
\end{figure}

To assess the numerical reliability of the time-domain evolution, as an example, 
we performed a convergence analysis for the tensor-coupled scalar
perturbation model with $\zeta=16$ (higher than critical values discussed in the next section), $\theta=0.2$,
$\mu=0.5$ and $\ell=2$, using three resolutions, $\Delta=0.3$, $0.15$ and $0.075$, with the same physical evolution time $t_{\max}=600$ at $r_*=0$.
The late-time exponential rate $\Gamma$ was extracted by a linear fitting
$\ln|\psi(t,r_*=0)|=\Gamma t+\mathrm{const}$ on the interval $t\in[350,550]$.
The convergence order was estimated from the successive differences of the
rates,
\begin{equation}
p=\log_2\left|
\frac{\Gamma_{0.3}-\Gamma_{0.15}}
{\Gamma_{0.15}-\Gamma_{0.075}}
\right| .
\end{equation}
For the initial pulse $\psi(u=0,v)=\exp[-(v-v_c)^2/2\sigma^2]$ with
$(v_c,\sigma)=(15,3)$, we obtain
\begin{equation}
\Gamma_{0.3}=0.0195623,\qquad
\Gamma_{0.15}=0.0196045,\qquad
\Gamma_{0.075}=0.0196151,
\end{equation}
where $p=2.00$ which is consistent with second-order convergence. The Richardson-extrapolated
rate is $\Gamma=0.019619$. The corresponding time-domain profiles are shown in the left panel of Fig.~(\ref{conv}). We also repeat our calculations for $r^*=24.9$ which is illustrated in the right panel. In order to examine the dependence of the behavior of ringdown profiles on the initial data, we considered the analysis for Gaussian pulses with
$( v_c,\sigma)=(10,2)$ and $(20,4)$ as well, but the extracted rates exhibit the same second-order convergence.

\subsection{Critical points and length scale for \(\theta\)}

In this subsection, we discuss the critical values of the parameter space
beyond which the perturbations undergo a transition from a stable ringing
phase to an unstable diverging phase, with particular emphasis on the coupling
constant $\zeta$ in the two scenarios. From this analysis we determine a length scale for the NC parameter $\theta$. For the remaining parameters we obtain the following
results. Increasing the scalar field mass $\mu$ enhances stability in both
models, so no critical point exists for $\mu$ in either model. For the
multipolar number $\ell$, the behaviour differs between the two models: in the
tensor model, increasing $\ell$ deepens the potential well, leading to a
critical value $\ell_{\text{c}}$ above which the system becomes unstable,
whereas in the scalar model, increasing $\ell$ raises the centrifugal barrier
and reduces the well depth, making the system more stable; hence no critical
point exists for $\ell$ in the scalar model. For the NC parameter $\theta$, as
shown in Fig.~(\ref{f7}), for each value of $\zeta$ an increase of $\theta$
leads to a critical value $\theta_{\text{c}}$ corresponding to an unstable
point.

The corresponding threshold values of the coupling constants from the ringdown
considerations for the two models are listed in Tab.~(\ref{T1}), in particular
for $\mu=0.5$ and $\mu=0$.
A striking difference between the two models is their opposite
dependence on the angular multipole number $\ell$.
In the scalar coupling scenario, $\zeta_{\mathrm{c}}$
increases monotonically with $\ell$, whereas in the tensor
coupling scenario it decreases, approaching progressively
smaller values at higher multipoles. Furthermore, introducing a nonzero scalar-field mass increases $\zeta_{\mathrm{c}}$ in both scenarios for all the considered
multipoles. This indicates that the mass term suppresses the onset of the instability, requiring a stronger coupling to reach the critical point. These findings are consistent with
the preceding discussion of the distinct instability mechanisms in the scalar and tensor coupling models and the stabilizing effect of the scalar-field mass. In the case of massless scalar fields our results listed in Tab.~(\ref{T1}) are also consistent with the values obtained in Ref.~\cite{Yan:2020nvk} for an NC geometry inspired black hole.

\begin{table}[H]
\centering
\caption{$\zeta_{\text{c}}$ of two scenarios when the test field is massive
or massless for $M/\sqrt{\theta}\simeq2.24$}
\label{T1}
\begin{tabular}{c|cc|cc}
\toprule
& \multicolumn{2}{c|}{$\mu=0.5$} & \multicolumn{2}{c}{$\mu=0$} \\
\cmidrule(lr){2-3} \cmidrule(lr){4-5}
$\ell$ & scalar model & tensor model & scalar model & tensor model  \\
\midrule
2   & 12.637  & 14.559 & 11.304  & 12.977 \\
4   & 28.123  & 9.052  & 27.016  & 8.696  \\
6   & 50.784  & 7.527  & 49.760  & 7.376  \\
10  & 117.416 & 6.424  & 116.461 & 6.373  \\
15  & 240.595 & 5.913  & 239.675 & 5.891  \\
20  & 408.077 & 5.668  & 407.175 & 5.656  \\
\bottomrule
\end{tabular}
\end{table}

Tab.~(\ref{T2}) presents the dependence of the critical coupling $\zeta_{\mathrm{c}}$ on the dimensionless parameter $M/\sqrt{\theta}$ in the two models. As $M/\sqrt{\theta}$ increases, the critical coupling grows rapidly in both scenarios which means that the onset of the instability requires increasingly stronger coupling. This growth becomes particularly pronounced at large values of $M/\sqrt{\theta}$, where
$\zeta_{\mathrm{c}}$ reaches the values very larger than its values at smaller ratios.

\begin{table}[H]
\centering
\caption{\(\zeta_{\text{c}}\) for different values of \(\theta\) when
\(\ell=2\), \(\mu=0.5\).}
\begin{tabular}{c|ccccccccc}
\toprule
\multirow{2}{*}{Model} & \multicolumn{9}{c}{\(M/\sqrt{\theta}\)} \\
\cmidrule(lr){2-10}
& 2 & 2.24 & 2.67 & 3.16 & 3.53 & 4.08 & 5 & 6.742 & 7.07 \\
\midrule
scalar model & 7.186 &12.636 & 51.550 & 427.571 & 3193.816 & 111385. & \(2.005\times 10^8\) & \(2.99\times 10^{17}\) & -- \\
Tensor model & 6.315 & 14.559 & 66.343 & 567.371 & 4278.031 & 150199. & \(2.714\times 10^8\) & \(1.42\times 10^{17}\) & -- \\
\bottomrule
\end{tabular}
\label{T2}
\end{table}

Before any physical interpretation of the values of $\zeta_{\text{c}}$ in the tensor model, we should examine the hyperbolicity condition for the effective kinetic metric presented in the equation of motion of the scalar field \cite{Brustein:2017iet}. In this respect, we can rewrite Eq.~(\ref{Eom1}) in the form 
\begin{equation}
\frac{1}{\sqrt{-g}}\,\partial_\mu\!\left[\sqrt{-g}\,\mathcal{G}^{\mu\nu}\,\partial_\nu\phi\right]
- \mu^2\phi = 0,
\qquad
\mathcal{G}^{\mu\nu} \equiv g^{\mu\nu} - \zeta\,G^{\mu\nu},
\end{equation}
where $\mathcal{G}^{\mu\nu}$ is called the effective kinetic metric. For the NC-Sch background given by (\ref{bm}), the non-vanishing
components of this kinetic metric are as follows:
\begin{equation}
\mathcal{G}_{\mu\nu}
=
\mathrm{diag}\!\Big(
-f(r)\big(1+\zeta K(r)\big),\;
\frac{1}{f(r)}\big(1+\zeta K(r)\big),\;
r^{2}\big(1-\zeta L(r)\big),\;
r^{2}\sin^{2}\vartheta\big(1-\zeta L(r)\big)
\Big),
\end{equation}
where $K(r)$ and $L(r)$ are defined in Eq.~(\ref{kl}). Using the blackening factor (\ref{Sch}) these functions become
\begin{equation}
K(r)=\frac{M}{\sqrt{\pi}\,\theta^{3/2}}\,e^{-r^{2}/4\theta},
\qquad
L(r)=K(r)\!\left(\frac{r^{2}}{4\theta}-1\right).
\end{equation}
The temporal and radial components of $\mathcal{G}_{\mu\nu}$ are proportional
to $1+\zeta K(r)$, which is manifestly positive for every $\zeta>0$ outside
the event horizon, thus no signature change arises from the radial sector.

The angular components, on the other hand, are proportional to
$1-\zeta L(r)$. Defining the hyperbolicity threshold as
\begin{equation}
\zeta_{\mathrm{hyp}}
\equiv
\left[\max_{r>r_{+}} L(r)\right]^{-1},
\label{zeta_hyp}
\end{equation}
then the requirement \(1-\zeta L(r)>0\) for all \(r>r_{+}\) imposes an upper bound on the values of the coupling parameter \(\zeta<\zeta_{\mathrm{hyp}}\). A numerical investigation reveals that, for all values of \(M/\sqrt{\theta}\) considered in Tab.~(\ref{T2}), as well as
for the tensor-model results reported earlier, the critical couplings \(\zeta_{\mathrm{c}}\) obtained in the tensor model exceed \(\zeta_{\mathrm{hyp}}\). Thus, the finite critical couplings
reported in this table are reached  after the condition required for hyperbolic evolution with respect to the background
time coordinate \(t\) has been violated. Specifically, at \(\zeta=\zeta_{\mathrm{hyp}}\), the factor \(1-\zeta L(r)\)
vanishes at some point outside the event horizon, rendering the principal tensor degenerate, while for \(\zeta>\zeta_{\mathrm{hyp}}\) this factor is a negative value, thereby obstructing the required hyperbolic time evolution. Consequently, although the onset of instability can be identified mathematically through the vanishing of the corresponding critical frequency, it cannot be interpreted as a physical dynamical instability of a well-posed evolution problem. Therefore, for the metric within the parameter range investigated, no physically admissible instability is identified in the tensor model considered here. It is worth mentioning that for the cases \(\mu \to 0\) or \(\ell \to \infty\), 
\(\zeta=\zeta_{\mathrm{hyp}}\) corresponds exactly to the degeneracy of the principal tensor and for \(\mu^2<0\) (a tachyonic field), one can obtain \(\zeta>\zeta_{\mathrm{hyp}}\), which requires further investigation and lies outside the scope of the current paper.

Here, an approximation for the model-dependent length scale is derived from the scalar sector of Tab.~(\ref{T2}). Note that the scalar sector is free of any hyperbolicity condition, since the coupling term \(\zeta R\phi^{2}\) is a mass-like term rather than a kinetic term, thus the effective kinetic metric remains \(g_{\mu\nu}\). The scalar sector therefore provides a clean basis for evaluating the scale associated with \(\theta\). Close to \(M/\sqrt{\theta} = 7.07\), it is evident that even such a large coupling cannot drive the system into an unstable regime: in this region, the perturbation becomes close to a stable state and the spectrum is dominated by long-lived modes.

As an interesting outcome of this analysis, the value \(M/\sqrt{\theta} \geq 7.07\) is employed to define a characteristic length scale for the NC parameter \(\theta\), yielding
\begin{equation}
\frac{M}{\sqrt{\theta}} \geq 7.07 \quad \Longrightarrow \quad
\sqrt{\theta} \lesssim \frac{M}{7.07}.
\label{ub}
\end{equation}
We emphasize that this relation should be regarded as a model-dependent estimate rather than a strict, universal bound, as it relies on the specific choice of the stability threshold within this framework.

To provide an illustrative numerical estimation for this length scale, we consider the mass of primordial black holes (PBHs). PBHs have a variety of implications for early-universe cosmology---formed from the collapse of large-amplitude density perturbations predicted in some inflationary models. These black holes are of special interest for several reasons: \textit{i}) they can be so light \(M_{\text{pbh}}\leq 10^{-18} M_{\odot}\) that Hawking radiation might conceivably be detected, \textit{ii}) the intermediate-mass region \(10^2 M_{\odot} \leq M_{\text{pbh}} \leq 10^6 M_{\odot}\) can provide the galactic dark matter, and \textit{iii}) the supermassive black holes with \(M_{\text{pbh}}\geq 10^6 M_{\odot}\) play a role in cluster dark matter or galactic centers \cite{Frampton:2015png}. There also exist investigations for constraints on the fraction of primordial black holes as dark matter with masses between \(10^{-19}-10^4 M_{\odot}\) in Ref.~\cite{Sato-Polito:2019hws}. A class of these black holes with masses around \(10^{-19} M_{\odot}\) will have evaporated by the present day due to Hawking radiation, and their abundance can be constrained by the effects of the emitted particles on the gamma-ray background. Therefore, they may survive at early-universe times when NC quantum effects are important. Substituting the mass of such very light black holes into Eq.~(\ref{ub}), an order-of-magnitude estimate for the length scale is obtained:
\begin{equation}
\sqrt{\theta} \lesssim 2.1 \times 10^{-17}~\text{m}.
\label{ub2}
\end{equation}

\section*{Conclusion}

In this paper, we carried out a comparative analysis of scalar field perturbations in the vicinity of NC-Sch black hole spacetimes, focusing on two distinct non-minimal curvature couplings: the scalar model (coupling with Ricci scalar $R$) and the tensor model (kinetic coupling to the Einstein tensor $G_{\mu\nu}$). Using the usual sixth-order WKB approximation and time-domain integration, we computed the QNMs and ringdown profiles, respectively. Then, we examined how the parameters $\theta$, $\zeta$, $\mu$, and $\ell$ affect the stability of the system. The main results of this comparison are as follows: Based on our calculations,  no physical instability was found in the tensor model within the NC-Sch spacetime. Furthermore, an opposite \(\ell\)-dependence of the threshold couplings was identified between the two models.

The results have shown that increasing $\theta$ or $\zeta$ suppresses the real part of the QNM frequencies, an effect that is negligible for the fundamental mode ($n = 0$) but becomes more pronounced at higher overtones. However, the imaginary part (damping rate) remains largely unaffected within the considered parameter ranges. Moreover, the scalar field mass $\mu$ influences only the real part, and this effect is already visible at the fundamental mode. For low overtone numbers, particularly the fundamental mode, the two models yield nearly identical QNM frequencies, indicating that the low-frequency dynamics is insensitive to the specific type of coupling.

From the ringdown profiles, we find that increasing $\theta$ or $\zeta$ renders both models unstable, increasing $\ell$ destabilises the tensor model but stabilises the scalar model, and increasing $\mu$ stabilises both models. In the scalar model, $\zeta_{\text{c}}$ increases monotonically with $\ell$ and it was shown that there is no physically admissible instability in the tensor model due to the hyperbolicity condition. The criticality was further examined for different values of \(M/\sqrt{\theta}\) 
(or equivalently, the NC parameter \(\theta\)). Based on the scalar sector, 
which is free of hyperbolicity issue, no instability was found for couplings 
up to \(\zeta \sim 10^{21}\) when \(M/\sqrt{\theta} \geq 7.07\). In this regime, 
the system behaves close to a stable state against scalar perturbations. This 
specific condition was then employed for very light primordial black holes to 
estimate an admissible, model-dependent length scale for the NC parameter, as 
given by Eq.~(\ref{ub2}). It should be emphasized that this value is a 
model-dependent estimate rather than a universal physical bound.

\section*{Acknowledgments}

The authors would like to thank V. Cardoso and M. Motahharfar for valuable comments and useful discussions. Specially, M.K. is grateful to them for discussions about programming and for sharing computational codes.


\end{document}